\documentclass[a4paper,11pt]{article}
\usepackage{jinstpub} 
\usepackage{lineno}
\usepackage{graphicx}
\usepackage{subcaption}
\usepackage{float}
\usepackage{placeins}
\usepackage{siunitx} 
\usepackage{textcomp}
\usepackage{hyperref}
\usepackage{soul}
\usepackage[toc]{appendix}
\usepackage{orcidlink}

\title{\boldmath Cryogenic UV detection using stress-engineered zero-bias ZnO-thin film based Piezo-Photonic detector}

\begingroup 
\footnotetext{These authors contributed equally to this work.}
\endgroup

\author{
P.~Sau\orcidlink{0000-0002-1755-9367}\textsuperscript{1,†},
N.~Hancock\textsuperscript{1,*},
I.~Tzoka\orcidlink{0000-0001-7811-5068}\textsuperscript{1,*},
V.~Khichar\textsuperscript{1}, 
A.~Barajas\textsuperscript{1}, 
G.~Gansle\textsuperscript{1}, 
N.~Hozhabri\textsuperscript{2}, 
V.A.~Chirayath\orcidlink{0000-0003-2450-2753}\textsuperscript{1}, 
J.~Asaadi\orcidlink{0000-0001-6915-5279}\textsuperscript{1}
}

\affiliation{\textsuperscript{1}Department of Physics, University of Texas at Arlington, Texas, USA 76019}
\affiliation{\textsuperscript{2}Nanotechnology Research Center, University of Texas at Arlington, Texas, USA 76019}

\emailAdd{pxs6807@mavs.uta.edu\,$^{\dagger}$}
\emailAdd{nicholas.hancock@uta.edu}
\emailAdd{iakovos.tzoka@uta.edu}
\emailAdd{jonathan.asaadi@uta.edu}

\abstract{We demonstrate a zero-bias ultraviolet (UV) detector using Zinc Oxide (ZnO) thin films as the active semiconductor layer, specifically for application in cryogenic conditions. The zero-bias device utilizes the piezoelectric potential developed through interfacial stress in the active semiconductor layer for charge transport. Here, we explored two vertically stacked Metal-Semiconductor-Metal (MSM) configurations: Sample I, a device comprised of Chromium (Cr)/ZnO/Cr layers, and Sample II, a ZnO–Silicon Nitride ($Si_{3}N_{4}$) device comprised of Cr/$Si_{3}N_{4}$/ZnO/Cr layers. The $Si_{3}N_{4}$ layer in Sample II was introduced in the form of pillars, with the aim of increasing the residual stress in the active region. These fabricated devices were tested in both room and cryogenic temperatures to characterize their UV-detection performance in a custom test-stand using a 365nm UV LED source. We observe a higher UV-induced voltage signal for Sample II in comparison to Sample I, both in room and cryogenic temperature measurements. Our Grazing-Incidence X-Ray Diffraction (GiXRD) measurements showed approximately 40\% higher residual stress in Sample II in comparison to that in Sample I. A higher residual stress possibly suggests a higher induced piezo-potential in Sample II in comparison to Sample I, which explains the enhancement in the UV-induced signal. Our results show that, through appropriate in-device stress engineering, UV photoinduced signals can be enhanced, increasing their sensitivity. A zero-bias photodetector with in-device stress engineering as demonstrated here can have applications in extreme environments, like cryogenic liquid noble elements or high radiation space environments, where low-or zero-power detection may be required.}


\begin{document}
\maketitle
\flushbottom

\section{Introduction}
\label{sec:intro}

Recently, the search for UV photodetectors has seen rapid progress due to its widespread applications that range from molecular fluorescence imaging, forensic analysis, flame detection, space sensors, ozone layer depletion monitors, and solar absorbers, as a few examples\cite{kumar2025recent,boruah2019zinc}. Various Metal-Semiconductor-Metal (MSM) detectors have been extensively explored for their utility in these UV-detector based applications~\cite{zhu2019recent}. However, most of these MSM detectors often require external biasing to produce efficient transport of the photo-induced charge carriers. These suffer from reduced performance in cryogenic and other extreme environments due to suppressed thermal carrier activation \cite{rahman2025carrier}. However, recent advances have shown that strain engineering in nanostructured ZnO can generate piezoelectric fields sufficient to drive carrier separation and enhance responsivity without external bias~\cite{willander2010zinc,ZHANG2014237}.This opens new possibilities for passive, low-power UV detection in extreme or bias-limited environments.

Zinc oxide (ZnO) is a wide-band gap semiconductor ($\sim$ 3.37 eV) with excellent optoelectronic and piezoelectric properties, making it a strong candidate for UV sensing applications in a variety of environmental conditions ~\cite{willander2010zinc, wang2021review, pandey2021review, coatings12111648}. ZnO's high exciton binding energy (60 meV), strong ultraviolet (UV) absorption, and transparency in the visible range support efficient UV photodetection. The noncentrosymmetric wurtzite crystal structure of ZnO allows for strain-induced polarization, giving rise to a piezoelectric potential that can assist in the separation of the photocarrier under zero bias conditions \cite{kandpal, ZHANG2014237}. Furthermore, the relatively low manufacturing cost, nontoxic nature, and relative ease of fabrication compared to similar solid-state detectors offer the opportunity to explore the enhancement of the induced residual stress due to interface strain engineering. 

In this work, we demonstrate a zero-bias ZnO-based UV piezophotonic detector, capable of operation under cryogenic conditions. Two sample geometries are studied: a planar ZnO thin film (referred to as Sample I), and a heterostructure (referred to as Sample II) incorporating Si\textsubscript{3}N\textsubscript{4} pillars designed to introduce localized interfacial stress, thereby changing the internal piezoelectric field within the ZnO layer. The thin films were fabricated on a double-side polished (DSP) intrinsic Silicon (Si) substrate using DC sputtering techniques. The uniformity, thickness and the crystalline sizes were characterized using cross-sectional and Field-emission Scanning Electron Microscope (FeSEM). This further enhances the reproducibility and thermal cycling of the fabricated devices throughout the entire device testing phase. 

Following the characterization of the fabricated layer's thicknesses, we present a comparative analysis of the UV-induced response of both devices at room temperature and at cryogenic conditions. Experimental results show that Sample II, with enhanced engineered interfacial stress, exhibits enhanced signal output under zero-bias operation, suggesting the role of piezoelectric enhancement in charge separation and transport. The light-induced tests in both room and cryogenic temperatures were performed using an in-house custom designed test stand. 

The residual stress present in the fabricated samples was characterized using Grazing-Incidence X-Ray Diffraction (GiXRD) analysis~\cite{ennaceri2019influence,welzel2005stress}. This testing protocol provides a non-destructive method for characterizing the stress in a material~\cite{chen2011residual}. The working principle of this method relies on maintaining a constant incident angle of the X-ray while varying the detector over a range of angles. This allows one to obtain the characteristic diffraction peaks which are broadened and/or shifted due to the presence of lattice strains. The shifts in the diffraction peaks are indicative of strain due to lattice mismatch. Additionally, the broadening of these peaks beyond that of the instrumental resolution is representative of microstructural distortions and broadening due to the crystalline size. The origin of these microstructural distortions can be accounted for by thermal stresses, and also growth procedures such as stage rotation, growth speed, throwing distance, etc. ~\cite{pr13041257}. The primary focus of this study is to compare the enhancement in the measured residual stress due to lattice mismatch from the introduction of Si\textsubscript{3}N\textsubscript{4} pillars in the device geometry. We observed an enhancement in the ZnO residual stress in Sample II in comparison to the residual stress in the ZnO thin film in Sample I, suggesting an increased induced piezo-potential. This can provide a suitable explanation for the enhanced UV-induced voltage signal that we observe in Sample II both in room temperature and cryogenic temperature measurements. Please note that we have not directly shown an increase in piezo  potential but our results provide indirect evidence like faster signal rise time and signal amplitude for Sample II in comparison to Sample I. We were able to repeat the enhancement on similar device geometries which gives us confidence that induced stress by the $Si_3N_4$ pillars and subsequent increase in piezo-potential  are responsible for the enhancement seen.   

These findings support the development of bias-free UV detectors optimized through stress engineering. Such detectors are well-suited for deployment in remote, cryogenic, or low-power environments, with potential applications in high-energy physics experiments, UV monitoring systems, and space-based platforms. Future exploration can be targeted towards exploring other device configurations, a more stable cryogenic testing system and also exploring application-specific testing. 

The sections that follow detail the fabrication processes for the samples including the characterization techniques used to quantify layer thicknesses and crystalline sizes. We briefly introduce the custom-built experimental setup for UV response testing. Experimental results from both room temperature and cryogenic temperature measurements are presented to elucidate the UV-induced performance of the detectors. The experimental results section also presents the measurement of the fabrication-induced residual stress in ZnO from the broadening of the characteristic GIXRD peaks, and we discuss them in relation to the UV photodetection enhancement. 

\section{Device Fabrication}

\begin{figure}[htbp]
\centering
\begin{subfigure}[t]{0.45\textwidth}
    \includegraphics[width=\textwidth]{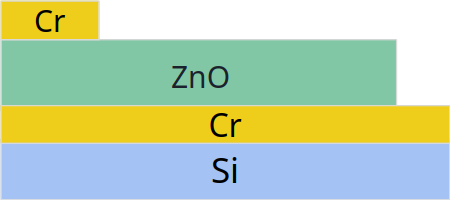}
    \caption{Cross-sectional schematic of Sample I.}
\end{subfigure}
\hfill
\begin{subfigure}[t]{0.45\textwidth}
    \includegraphics[width=\textwidth]{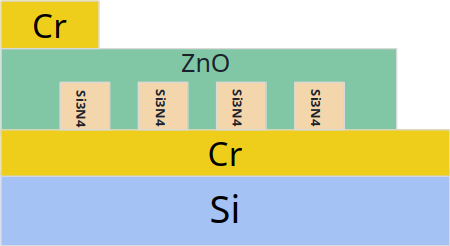}
    \caption{Cross-sectional schematic of Sample II.}
\end{subfigure}

\vspace{0.5em} 

\begin{subfigure}[t]{0.45\textwidth}
    \includegraphics[width=\textwidth]{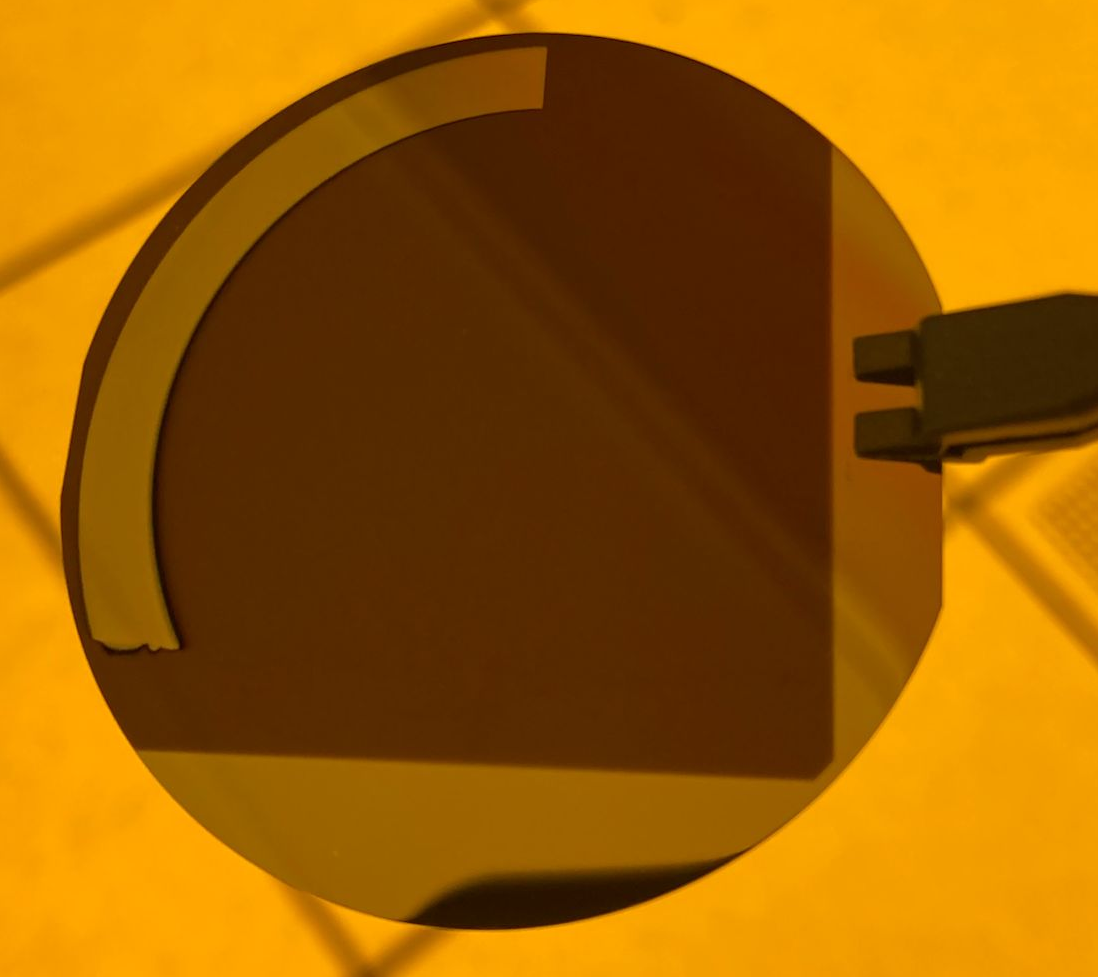}
    \caption{Fabricated Sample I.}
\end{subfigure}
\hfill
\begin{subfigure}[t]{0.45\textwidth}
    \includegraphics[width=\textwidth]{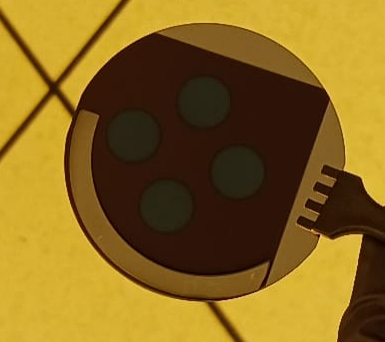}
    \caption{Fabricated Sample II.}
\end{subfigure}

\caption{(a), (b) Schematic showing the cross-sectional representation of a semi-vertical planar ZnO thin film (Sample I) and a semi-vertical heterostructure incorporating Si\textsubscript{3}N\textsubscript{4} pillars designed to introduce localized interfacial stress (Sample II) (note: the diagrams are not to scale) grown on an intrinsic Si substrate. For the purposes of this study, four pillars of Si\textsubscript{3}N\textsubscript{4} are grown underneath the layer of ZnO to induce in-device stress due to lattice mismatch. (c), (d) show images of the actual fabricated devices.}
\label{fig:devicefab}
\end{figure}

\FloatBarrier

Both samples were fabricated in the Shimadzu Institute for Nanotechnology Research Center (NRC) at UTA \cite{Shimadzu_NanoCenter}, inside a class-100 cleanroom. The thin-films were deposited using plasma sputtering systems. 

The ZnO devices were fabricated on a 2-inch, 500 ± 25$\mu$m thick, double-side-polished intrinsic Si wafer~\cite{WaferPro2024}. The intrinsic silicon wafers were cleaned in a piranha solution to remove any organic contaminants. Following that, an approximately 1$\mu$m thick chromium (Cr) layer was deposited using a Plasma Sputter operated at 150~W DC power with a strike gas (Argon) pressure of 40~mT. For Sample I, a $\sim50$ nm thick thin film of ZnO was deposited using a multi-gun AJA RF sputtering system \cite{AJA_Orion_Series} at a DC power of 100~W, with an Argon pressure of 4~mTorr. A shadow mask was used to deposit the ZnO thin layer within the intended region on the silicon substrate. This was done so that a small region of the bottom metal electrode remained exposed so that the electrical contacts could be made to the fabricated devices for performing light-induced electrical testing. Following the deposition of ZnO, the top metal contact of 1$\mu$m of chromium was deposited. Another custom designed shadow mask was used for the geometry of the top Cr layer, to coincide with the metal contact region on the printed circuit board in the experimental setup. For the case of Sample II, pillars of around 25 nm thick Si\textsubscript{3}N\textsubscript{4} and 1mm in diameter, were deposited on the bottom Cr layer using the AJA sputtering tool operated at 150~W with an Argon pressure of 5~mTorr. A hard shadow mask was used to create a 4-pillar pattern of diameter 1mm for the Si\textsubscript{3}N\textsubscript{4} layer. After this, the layers of ZnO and Cr were grown in the same way as previously mentioned for the Sample I fabrication procedure. Table \ref{tab:samples} summarizes the intended thickness for the different layers for Samples I and II. The fabricated thicknesses were later characterized using cross-sectional SEM. The measurements revealed that the actual layer thickness were within 5\% tolerance of our intended thicknesses. Figure~\ref{fig:devicefab} show the schematics and the fabricated images for Sample I and Sample II.

\begin{table}[htb]
\centering
\begin{tabular}{l|c|c|c}
\hline
Sample & Metal Contacts [Cr] & ZnO Thickness & Si$_3$N$_4$ Thickness \\
\hline
Sample I & 1\,µm & 50\,nm & - \\
Sample II & 1\,µm & 50\,nm & 25\,nm \\
\hline
\end{tabular}
\smallskip
\caption{Summary of layer composition and  intended thicknesses for the two fabricated samples. 
\label{tab:samples}}
\end{table}

\begin{figure}[htb]
\centering

\begin{subfigure}[t]{0.48\textwidth}
    \includegraphics[width=\textwidth]
    {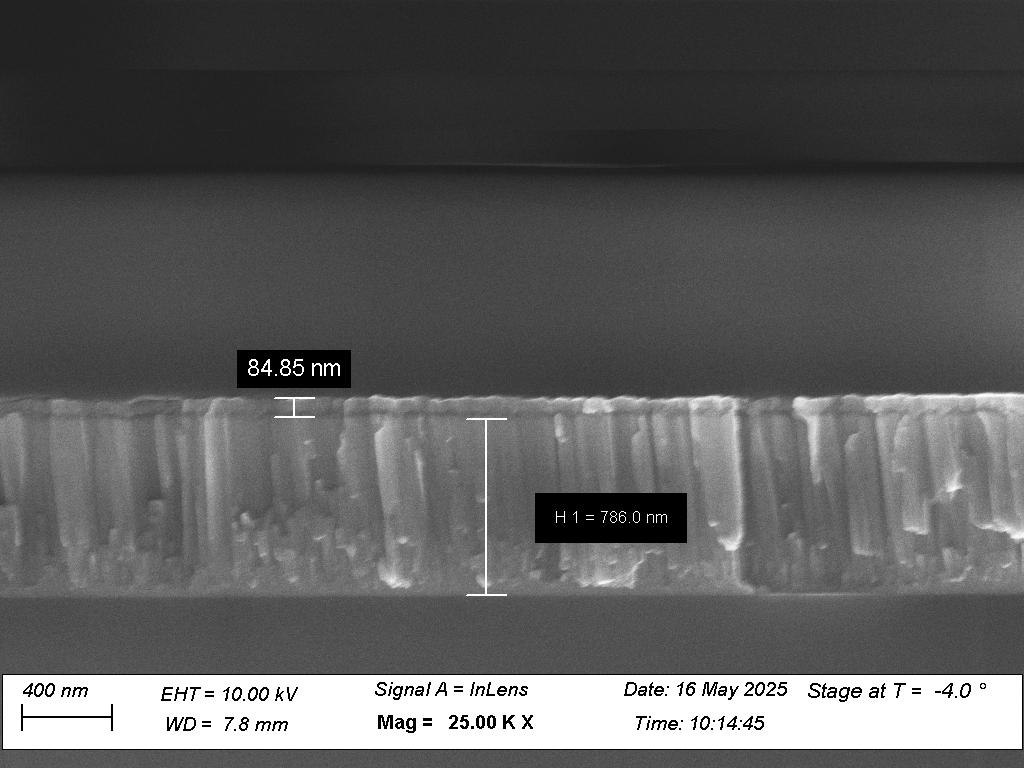}
    \caption{Sample II overall stack.}
    
\end{subfigure}
\hfill
\begin{subfigure}[t]{0.48\textwidth}
    \includegraphics[width=\textwidth]
    {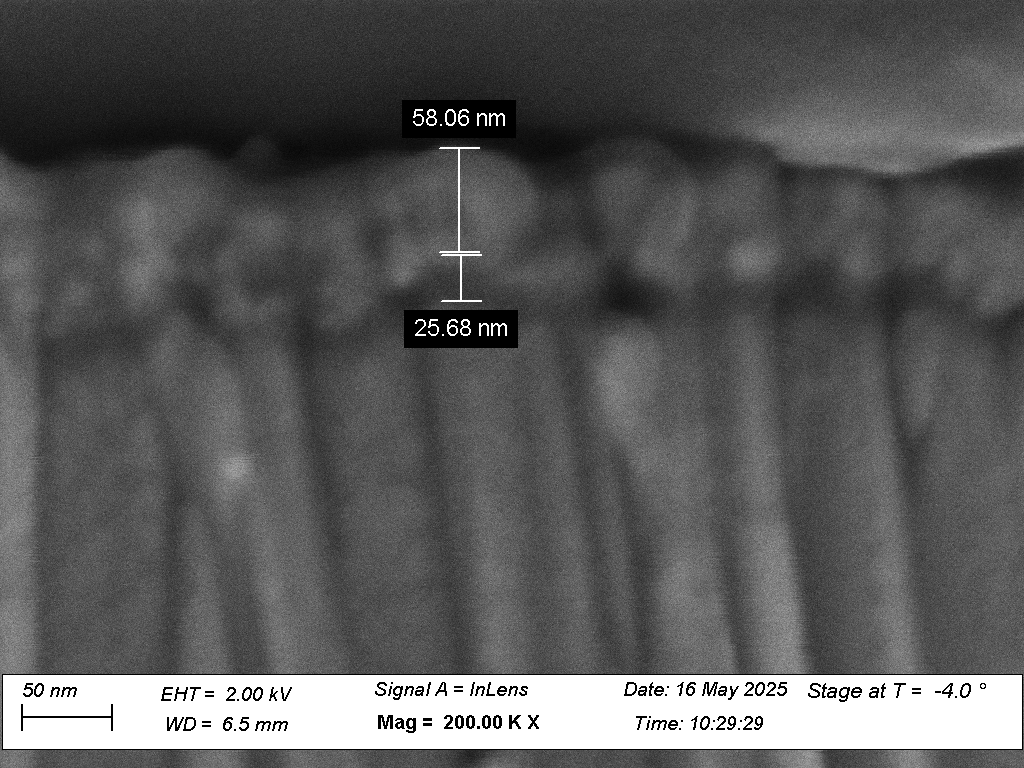}
    \caption{Sample II ZnO-Si\textsubscript{3}N\textsubscript{4} bilayer.}

\end{subfigure}

\vspace{0.5cm}

\begin{subfigure}[t]{0.48\textwidth}
    \includegraphics[width=\textwidth]{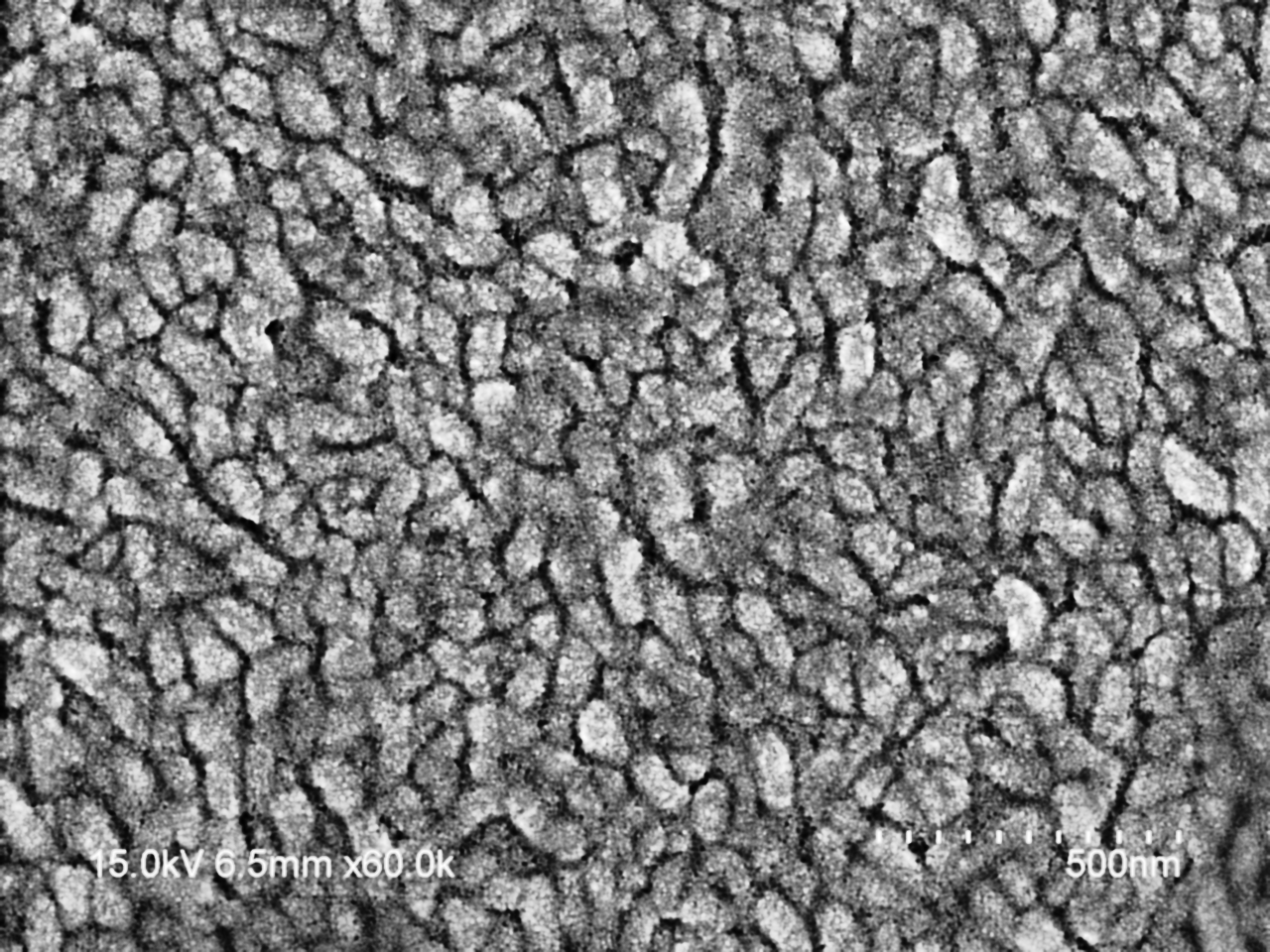}
    \caption{Sample I FeSEM image.}
\end{subfigure}
\hfill
\begin{subfigure}[t]{0.48\textwidth}
    \includegraphics[width=\textwidth]{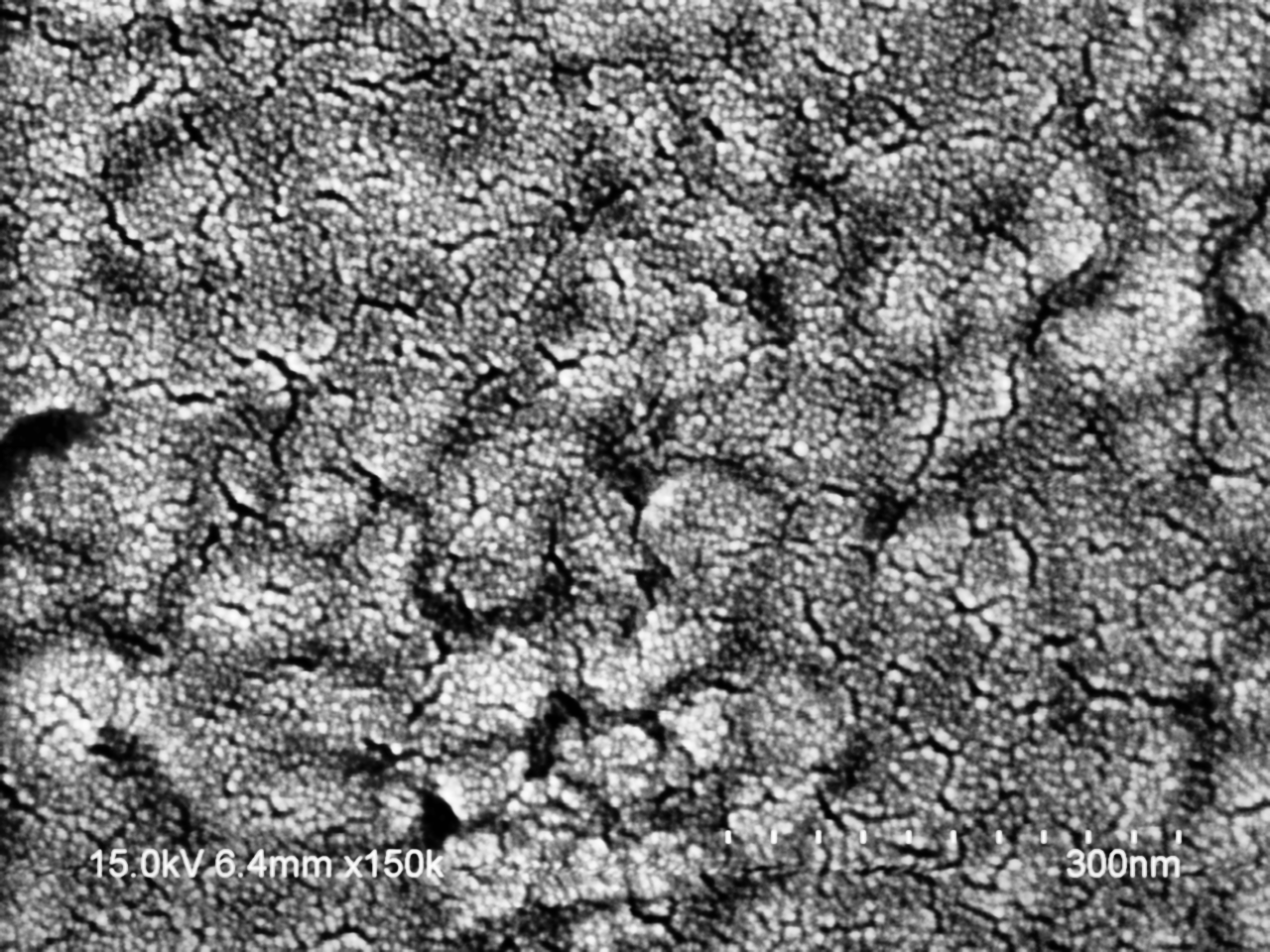}
    \caption{Sample II FeSEM image.}
\end{subfigure}

\caption{The cross-sectional SEM images of the ZnO–Si\textsubscript{3}N\textsubscript{4}/Cr/Si stack are shown in (a) and (b). (a) The overall stack shows the bottom Cr layer and the ZnO–Si\textsubscript{3}N\textsubscript{4} bilayer. (b) A zoomed-in view resolves the distinct ZnO and Si\textsubscript{3}N\textsubscript{4} layers. (c) and (d) show representative FeSEM scans for Samples I and I,I respectively. }
\label{fig:csSEMfigures}
\end{figure}

Figure \ref{fig:csSEMfigures}a shows the low-magnification cross-sectional Scanning Electron Microscope (SEM) image of Sample II, which shows the thickness of each uniformly deposited layer. Figure \ref{fig:csSEMfigures}b shows a higher magnification cross-sectional SEM plot of this device, which shows the interface between the ZnO and the Si\textsubscript{3}N\textsubscript{4} layer. We observe no particular delamination, and the film integrity appears intact.

Figure \ref{fig:csSEMfigures} (c) and (d) show FeSEM scans that were performed at the UTA Characterization Center for Materials and Biology \cite{UTA_CCMB} on representative samples for Sample I and II. These images show the crystalline sizes of ZnO near the surface of the deposited ZnO layer. These images were then analyzed using the ImageJ software \cite{ImageJ} to determine the ZnO crystalline sizes.


\section{Experimental Setup}\label{sec:ExperimentSetup}

An in-house, custom designed and built cryogenic test stand was used to test the room temperature and cryogenic UV response of these fabricated photodetectors. The details of the design and assembly of this test stand are similar to the one previously used in \cite{rooks2023development}. Figure~\ref{fig:cryoteststand} shows the 3D cross section of the CAD model of the test stand. Minor modifications to the cooling copper block and mounting structure for the samples, as well as the addition of an internal UV fiber to better direct the incident light, constitute the major changes to the setup. 

\begin{figure}[htb]
\centering
\includegraphics[width=.6\textwidth]{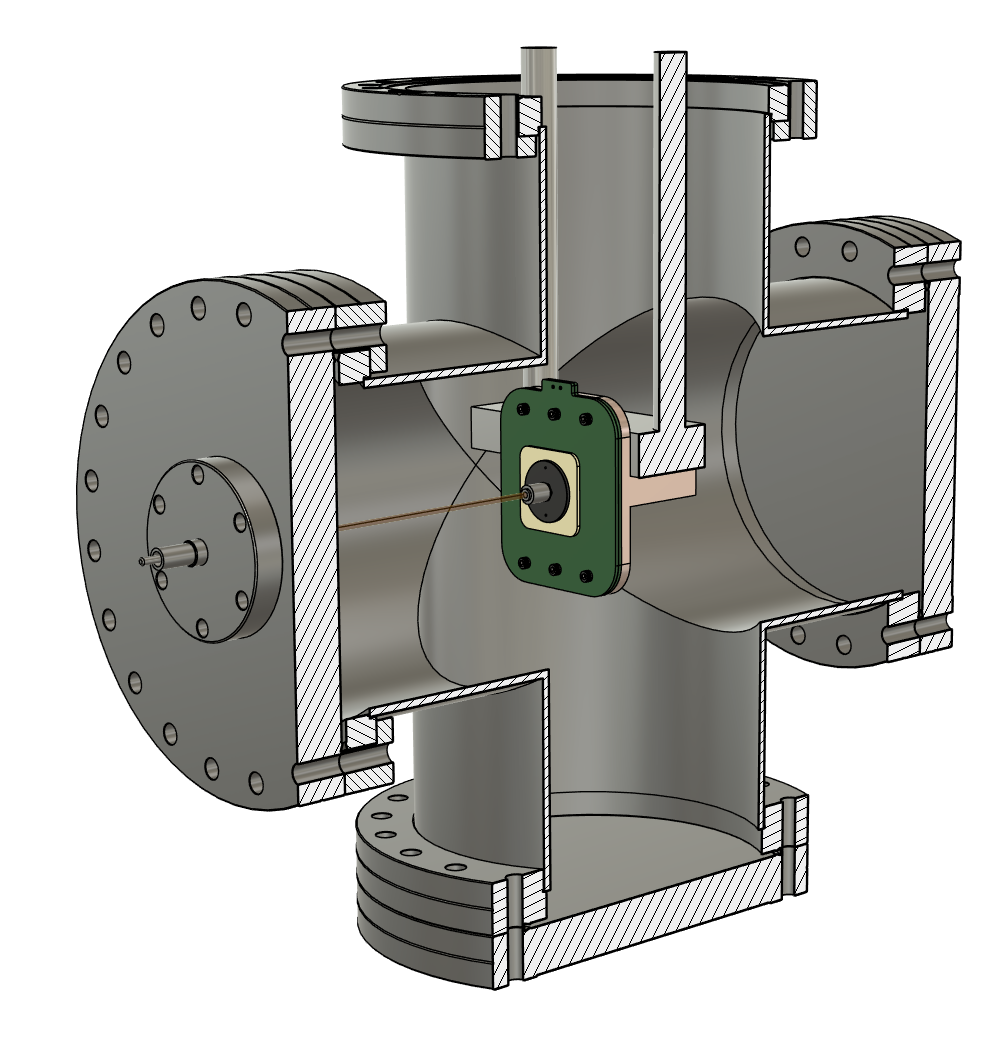}
\caption{3D cross-sectional CAD image of the in-house, custom-designed experimental setup operating under cryogenic temperatures and vacuum conditions. The setup is employed for the UV-induced testing of the fabricated devices. \label{fig:i}}
\end{figure}
\label{fig:cryoteststand}

\begin{figure}[htb]
\centering
\begin{subfigure}[t]{0.45\textwidth}
    \includegraphics[width=\textwidth]{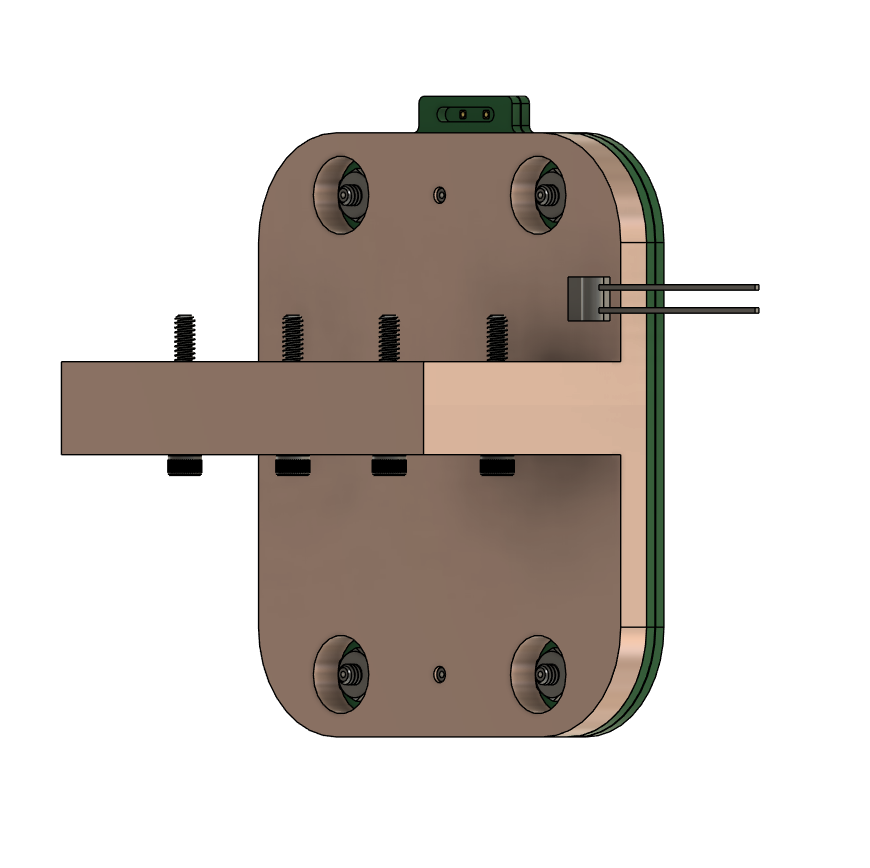}
    \caption{RTD behind the copper block}
\end{subfigure}
\hfill
\begin{subfigure}[t]{0.45\textwidth}
    \includegraphics[width=\textwidth]{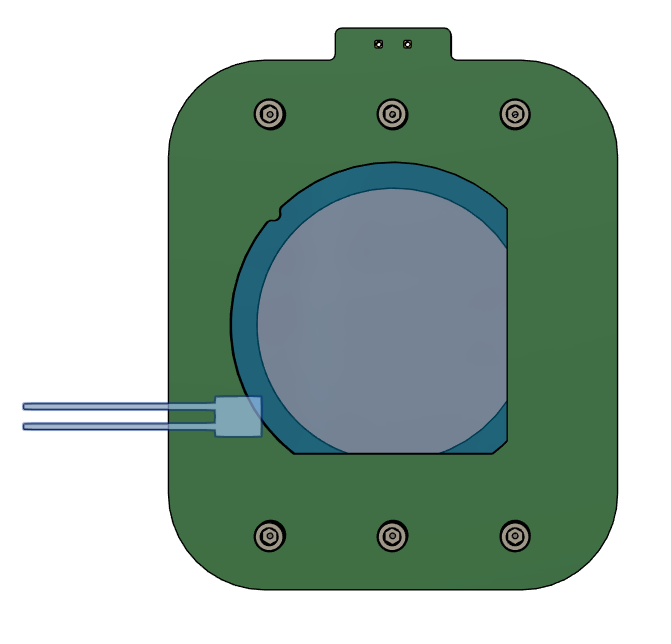}
    \caption{RTD in front of the copper block but behind the PCB sandwich}
\end{subfigure}
\hfill
\begin{subfigure}[t]{0.45\textwidth}
    \includegraphics[width=\textwidth]{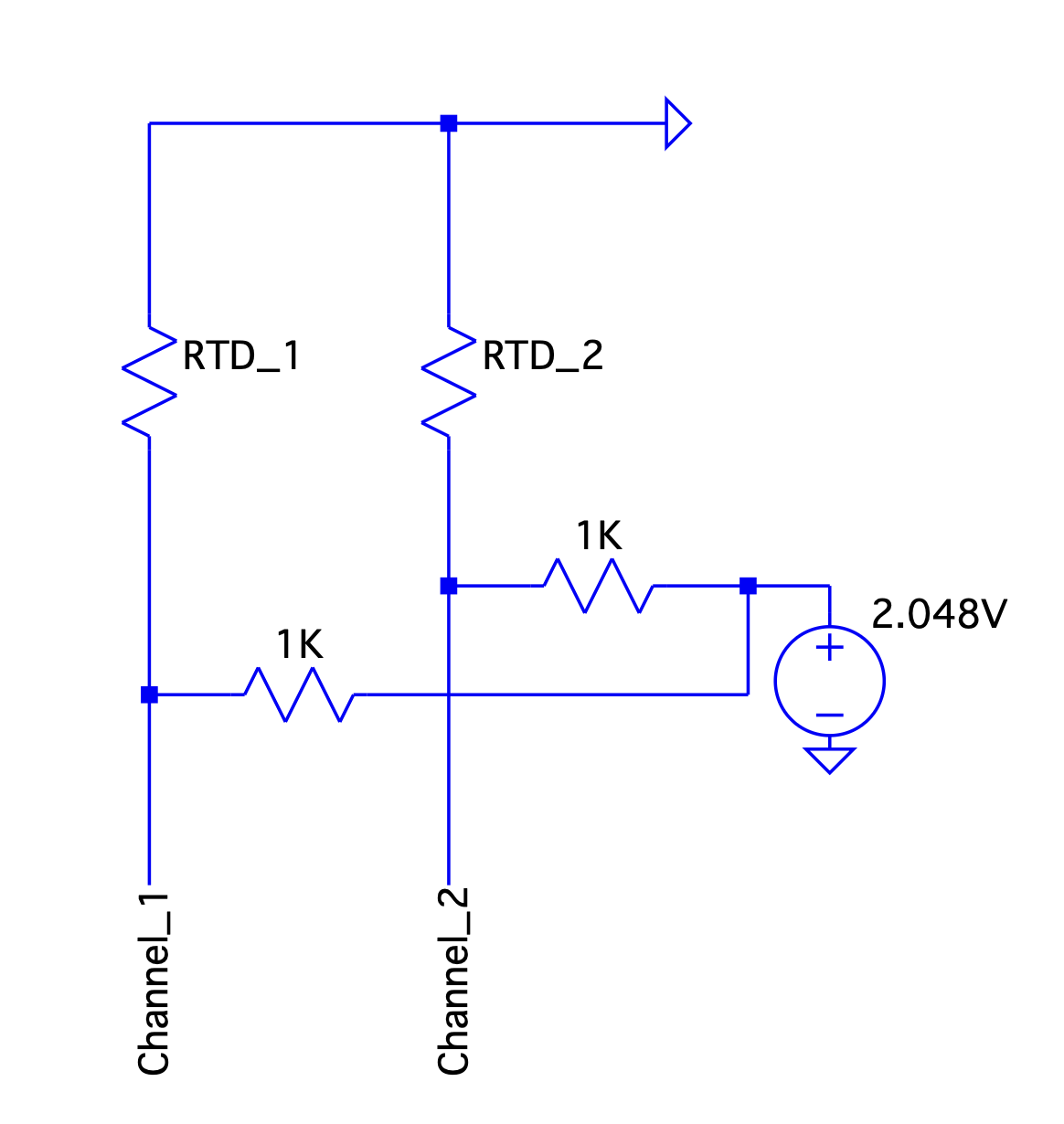}
    \caption{RTD Circuit Schematic}
\end{subfigure}
\caption{3D model showing the positions of the three RTDs used to calibrate the temperature during a cryogenic cycle: (a) behind the copper block, (b) in front of the copper but behind the PCB, and (c) LT Spice schematic of the RTD readout circuit}
\label{fig:RTDCAD}
\end{figure}

The entire chamber is pumped down to vacuum ($\sim$ 20~mTorr). The devices are mounted in a mechanical contact between two in-house custom-designed printed circuit boards (PCBs) to establish electrical contact to the top and bottom Cr electrodes in a semi-vertical structure. The PCBs along with the device are then mounted onto a copper block using vented socket head screws to reduce out-gassing. This copper block is mounted onto a stainless-steel bar, through which liquid nitrogen can flow, that cools the copper block, which then cools the device. To calibrate the rate of cooling and quantify the temperature gradient between the copper block and the cold finger, a temperature test was performed using NB-PTCO-381 platinum thin-film Resistance Temperature Detectors (RTDs)~\cite{TE_NBPTCO381}. Two RTDs were placed – one behind the copper blocks and one in front of the copper block (i.e. behind the sample) as shown in Figure~\ref{fig:RTDCAD}. The resistances of the RTDs were measured using a 16-channel ARD-LTC2499 24-bit ADC breakout board connected on top of an Arduino Uno. This allows for the measured resistances to be converted to temperature readings. The schematic for the RTD connections and the corresponding readout scheme is shown in Figure \ref{fig:RTDCAD} (c). The rate of flow of liquid Nitrogen was controlled using a manual valve connected to the LN2 dewar. For all the measurements done for the purposes of this paper, the manual valve was kept open at a constant position, to maintain the consistency of the cooling between each run.

\begin{figure}[htb]
\centering
\includegraphics[width=.8\textwidth]{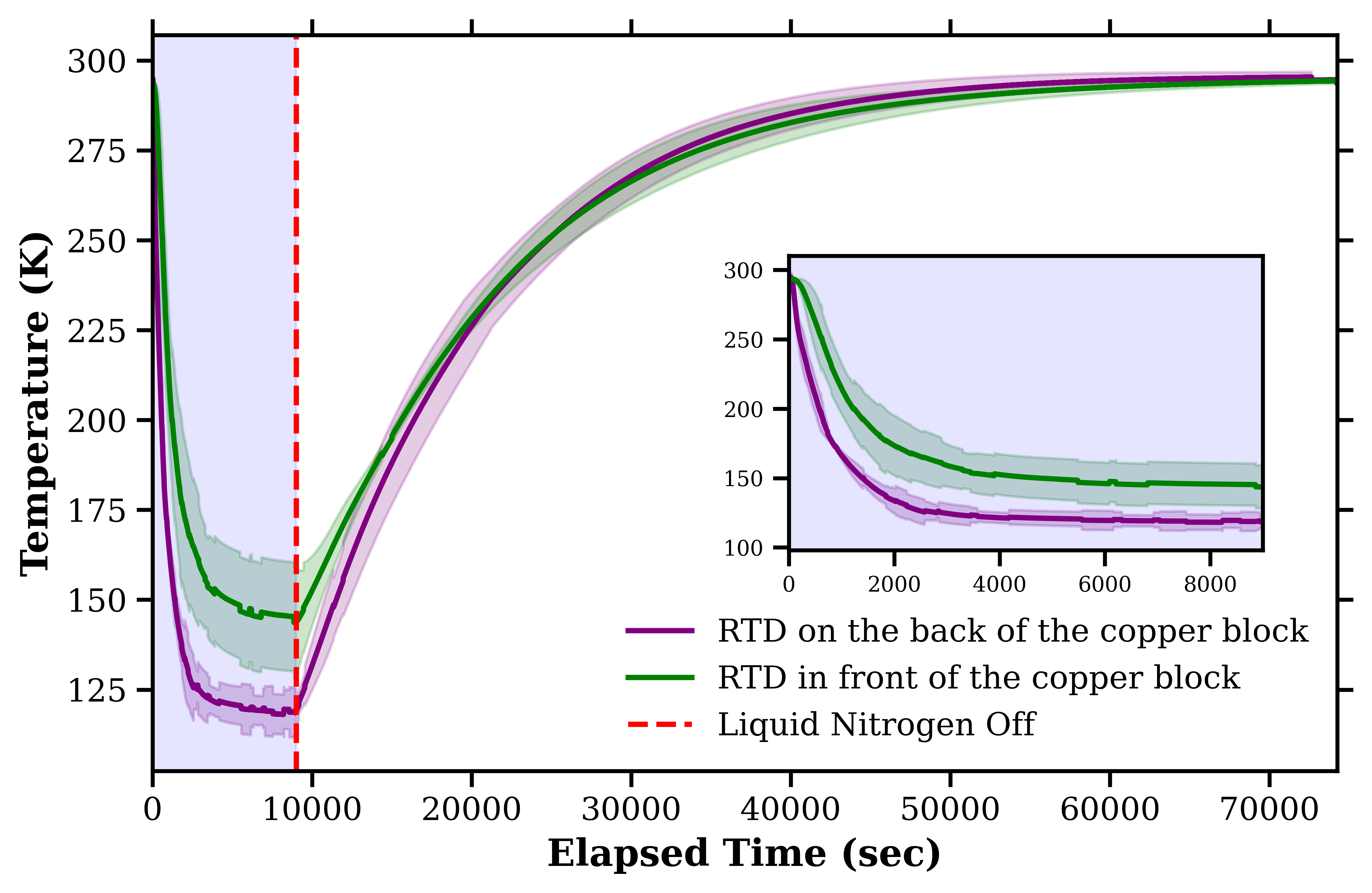}
\caption{The temperature calibration curve is shown for the RTDs placed behind the copper block and another infront of it. \label{fig:tempcalibrationCu}}
\end{figure}

Figure~\ref{fig:tempcalibrationCu} shows the temperature curve as a function of elapsed time for the two RTDs. For these temperature curves, liquid nitrogen was flowed through the cold finger for approximately 2.5 hours, then closed, and the system was allowed to warm back to room temperature.  We observe a temperature gradient of approximately 20 K between the back of the copper block and the front of the copper block (or behind the sample). Additionally, we assume that the temperature of the sample would be the same as the temperature on the front of the copper block with negligible deviations due to the minimal thickness of the devices (micrometer scale). We observe that this gradient quickly decreases and an equilibrium is reached between the 2 RTDs as they approach back to room temperature. Due to a slightly higher temperature gradient and the rapid decrease in the temperature during the cooling period, we performed the cryo-messaging during the cooling period. This allowed us to perform multiple short-lived UV-induced runs. The duration of each run lasted approximately 5 minutes. The small temperature variation during the data collection process was considered during the data analysis (\ref{appendix:Data Analysis}) and is represented as error bars on the horizontal axis in the plots shown later. These calibration curves were used to calibrate the low-temperature experimental data as mentioned later in Section \ref{sec:lowTempMeasurement}. 

The experimental setup and readout scheme used to perform this experiment includes a Thorlabs M365FP1 365nm UV LED light source \cite{Thorlabs_M365FP1} 
,which was connected to one of the ends of the test stand using a VUV-compatible $1000\mu$m optical fiber \cite{1000UV-Air}. This optical fiber was coupled to another vacuum compatible $100\mu$m optical fiber \cite{100UH-UV} at the feedthrough flange inside the chamber, which connects onto the custom designed connector placed on top of the PCB as shown in Figure~\ref{fig:cryoteststand}. The light intensity and frequency of flashing were controlled using a High Power Thorlabs LED Driver \cite{Thorlabs_DC2200}. For measuring the voltage signal, a coaxial cable was connected to the metal pins of the PCB board. This cable was then attached to a feedthrough flange, to which the connections of the Keithley 2182A nanovoltmeter~\cite{tek2182a_2025} were made using a spliced SHV cable.
Additionally, to avoid the light from the LED directly hitting the exposed metal contacts, and thus causing an induced photoelectric signal, a custom-made mask was designed to ensure that only a small active region of the ZnO was exposed to the incident light.

\section{Experimental Results}

\subsection{Room Temperature measurements of light-induced UV response}

For obtaining room temperature measurements, as discussed in Section \ref{sec:ExperimentSetup}, the samples are loaded in the test stand. The chamber is pumped down to a vacuum of around 20 mTorr. The 365~nm UV LED is used to shine light onto the samples, and the UV-induced signal is collected using a Nanovoltmeter. The Nanovoltmeter is set at a rate of 1 PLC \cite{Keithley_2182A} to reduce the presence of noise in the measured signal. To confirm that any observed signal is due to the response of the sample under test, an identical run is made but the light from the UV LED is blocked. No signal was observed during these controlled runs, which provides confidence that observed voltages are in fact due to UV induced signals.

\begin{figure}[H]
\centering
\begin{subfigure}[t]{0.62\textwidth}
    \centering
    \includegraphics[width=\textwidth]{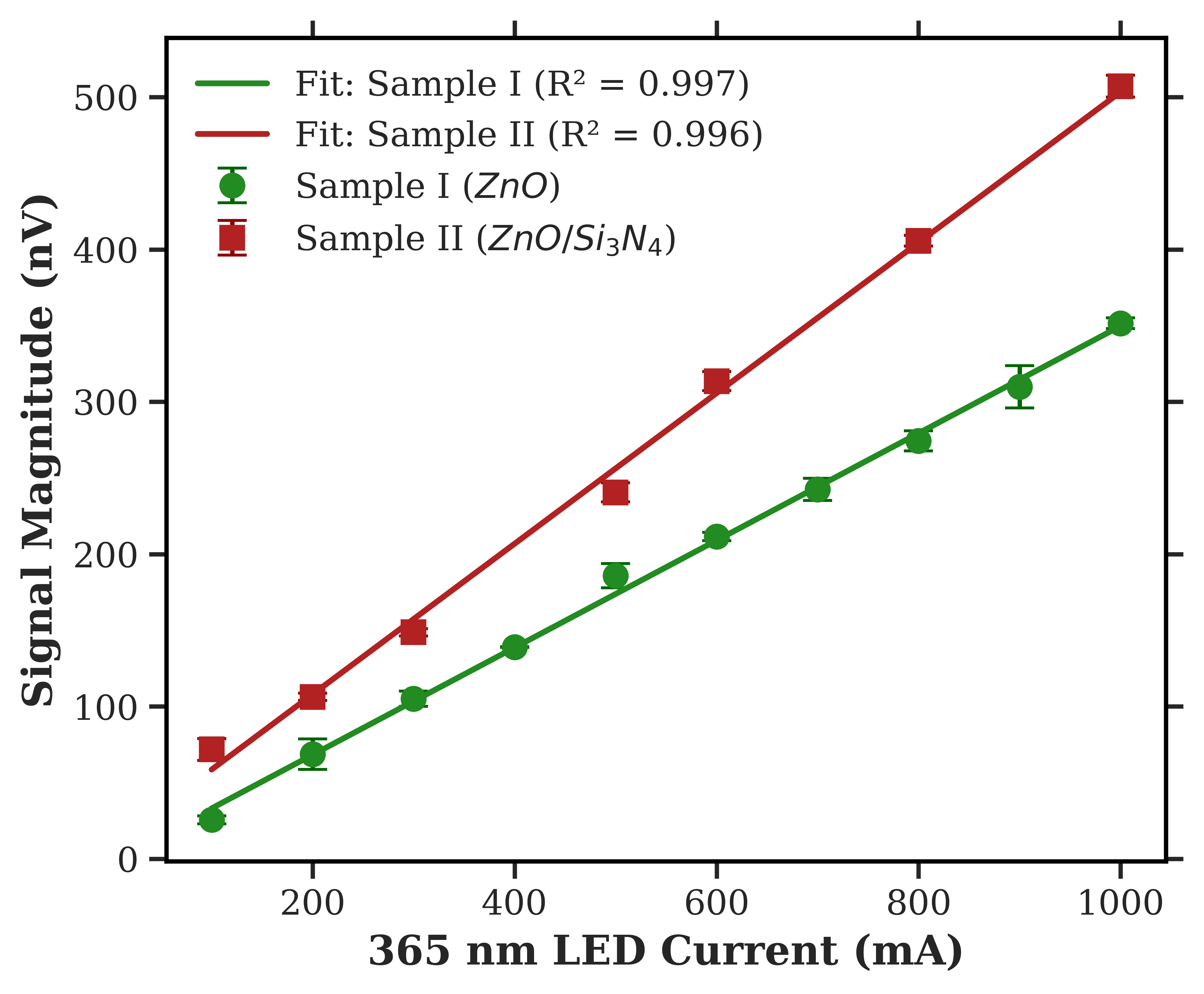}
\end{subfigure}
\hspace{0.1cm}
\begin{subfigure}[t]{0.34\textwidth}
    \centering
    \includegraphics[width=\textwidth]{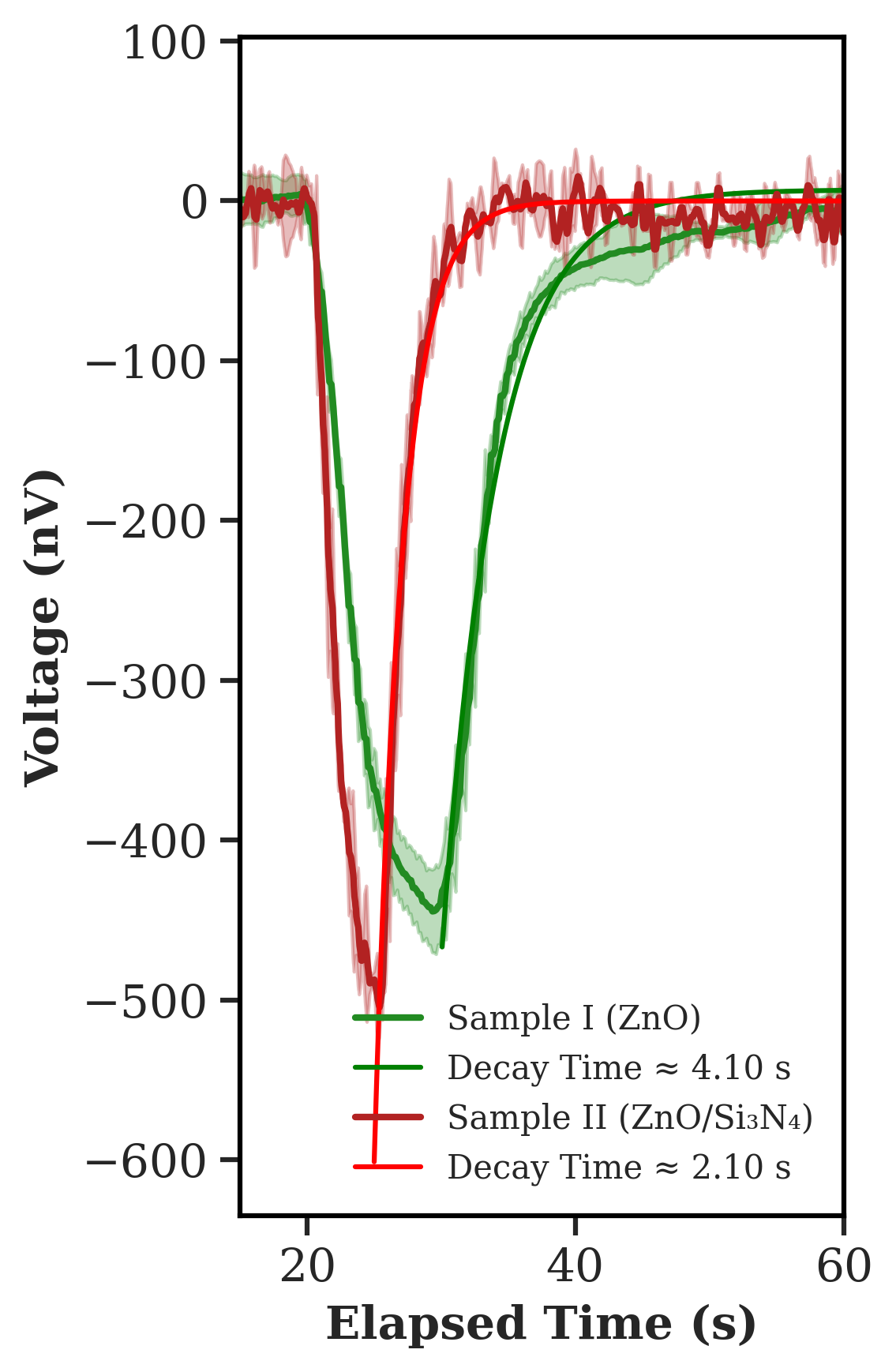}
\end{subfigure}
\caption{(a) Show the magnitude of the 365 nm UV-induced voltage signal as a function of LED current for the planar ZnO  sensor (Sample I–green) and the ZnO/Si$_3$N$_4$ (Sample II–red). An enhancement of $\sim 50\%$ due to strain engineering can be observed. (b) Room-temperature UV-induced voltage response at 1000 mA LED current for Sample I and Sample II. For Sample I, the UV light was applied for 10 s; however, to enable a consistent comparison between the two samples, the signal magnitude was evaluated over the first 5 s following UV illumination.}
\label{fig:roomdata}
\end{figure}

For measuring the UV photo response of the device at room temperature, a singular pulse of the LED with an ON-time of 5 s was used. This was done at various LED currents. To measure if there is any hysteresis effect, the spectra were collected both while decreasing the LED current (downscan), and then again during the increasing current (upscan) run as well. We observed no significant hysteresis effect in these signals. Figure \ref{fig:roomdata} is obtained by taking the magnitude of the peak of the UV-induced signal corresponding to the baseline of the signal when the UV was off. The scatter points represent the average magnitude of the UV-induced signal between the upscan and downscan runs. The error bars are calculated by taking the absolute value of the difference between the peak magnitude of the signal during the two UV-intensity-scan runs.

We observe an approximately $50 \%$ increase in the magnitude of the UV-induced signal for Sample II compared to Sample I. Additionally, we notice that the rate of increase of the magnitude of the signal varies between the two samples. We hypothesize that this is due to the varying quantum efficiency of the produced UV-induced charge, and also the varying transport efficiencies of the charges to their respective electrodes. This variation in the transport properties is also evident from the varying rise and decay times of the UV-induced signals. We fit an exponential function to the decaying edge of the signal. From this fit to the data, we obtain a decay time constant of approximately 4.10~s for Sample I and 2.10~s for Sample II (Figure~\ref{fig:roomdata}). Furthermore, the rise times (defined as the time required for the signal to increase from $10\%$ of the peak to $90\%$ of the peak value) for Sample I and II are 3.37s and 2.87s respectively. 

\subsection{Low Temperature measurements of light-induced UV response}\label{sec:lowTempMeasurement}

\begin{figure}[H]
\centering
\begin{subfigure}[c]{\textwidth}
    \includegraphics[width=\textwidth]{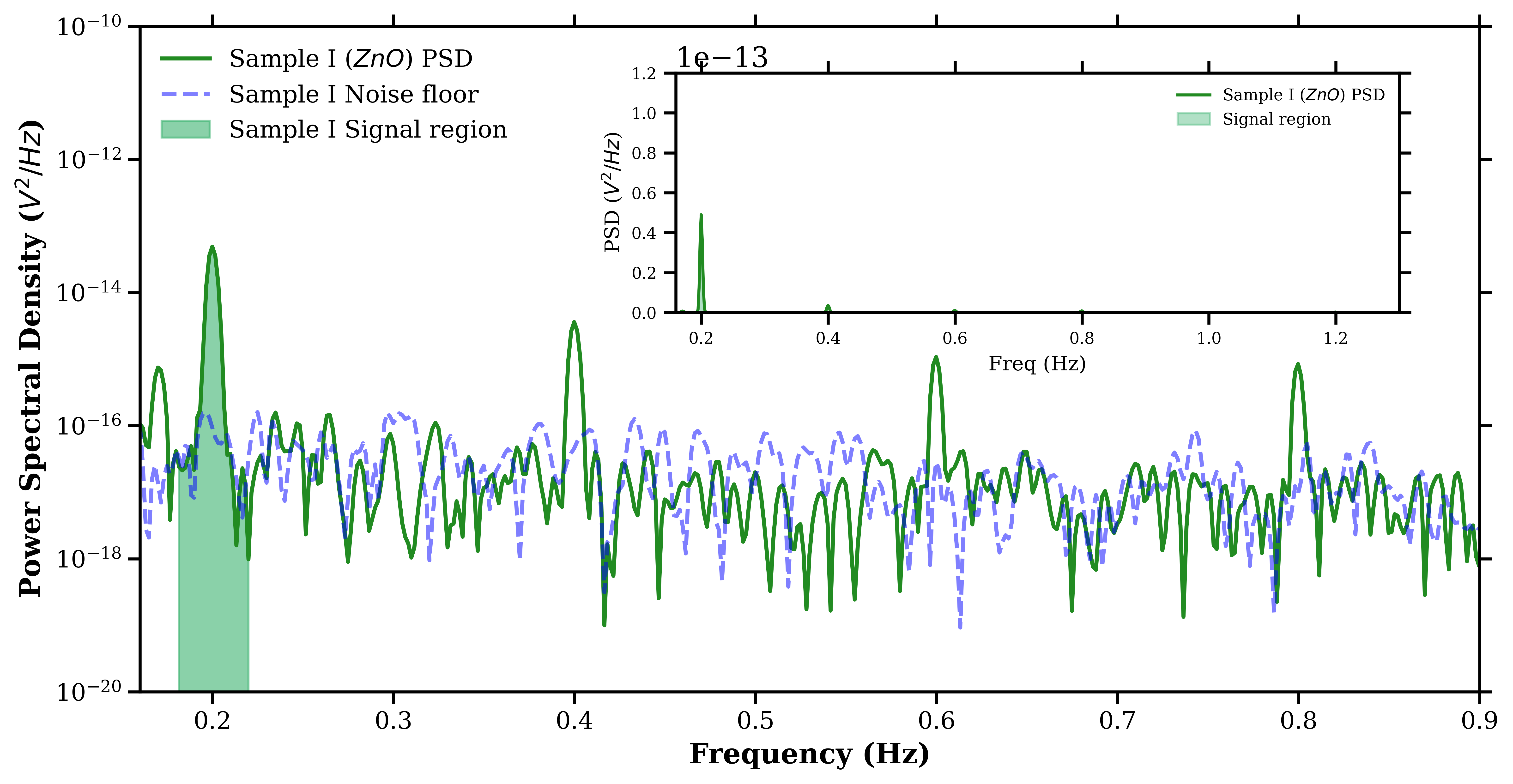}
\end{subfigure}

\begin{subfigure}[c]{1.0\textwidth}
    \includegraphics[width=\textwidth]{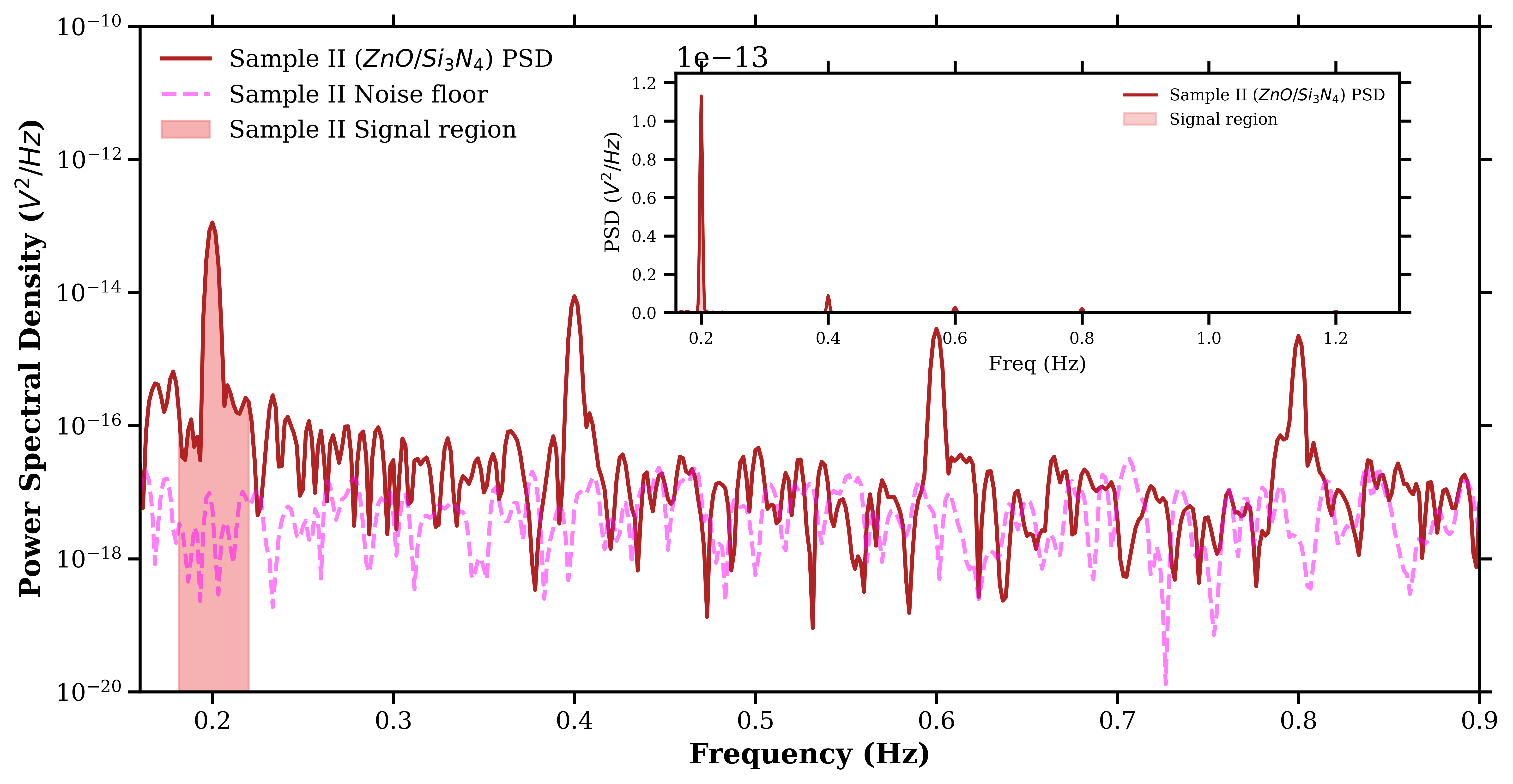}
\end{subfigure}
\caption{The solid lines show the frequency-domain voltage signals received from Samples I and II at a temperature of 164~$\pm$~6~K. The dashed lines show the signal recorded during the background run for each sample. The total area in the shaded region ($A_{sig}$) is the value used to calculate the SNR. The inset plots illustrate that the vast majority of the power of the device response is contained in the first peak and higher harmonics are not considered. The linear scale also makes the enhancement due to the $Si_3N_4$ pillars more visible.}
\label{fig:cryo_data_frequency_domain}
\end{figure}

To obtain the measurements at low temperatures, first, the sample in the test stand is cooled by flowing liquid Nitrogen through the cold finger for approximately 2.5 hours (similar to the temperature calibration runs mentioned previously in Figure \ref{fig:tempcalibrationCu}). Following this, liquid nitrogen is turned off and a background spectrum is collected for 5 minutes, during which the UV LED is pulsed 60 times at 1000mA, with an ON time of 2~s and an OFF time of 3~s, but the light is blocked from reaching the sample. Following this background spectrum, the first signal measurement is taken by pulsing the LED in the same way but with the light no longer blocked. Subsequently, until the temperature of the sample reaches nearly room temperature, 19 more such measurements are taken. These time domain spectra are then analyzed in the frequency space as described in Appendix \ref{appendix:Data Analysis}, to obtain the Power Spectral Density (PSD), which represents the power output per unit frequency times the impedance of the device. Figure \ref{fig:cryo_data_frequency_domain} shows the PSD for a representative low-temperature run for both samples. The peaks at multiples of 0.2~Hz provide evidence of the device responding to the 0.2~Hz pulsing of the UV-LED. These peaks are broadened because of the multiplication by the Hann window function in the time domain during the analysis. Therefore, to quantify the magnitude of the response, we calculated and compared the total area under the fundamental frequency (0.2~Hz) with the area under the background measurement in the same region of frequency space. Higher harmonics were not considered because the majority ($\gtrsim$80\%) of the power is contained within the first peak. 

\begin{figure}[H]
\centering
\includegraphics[width=\textwidth]{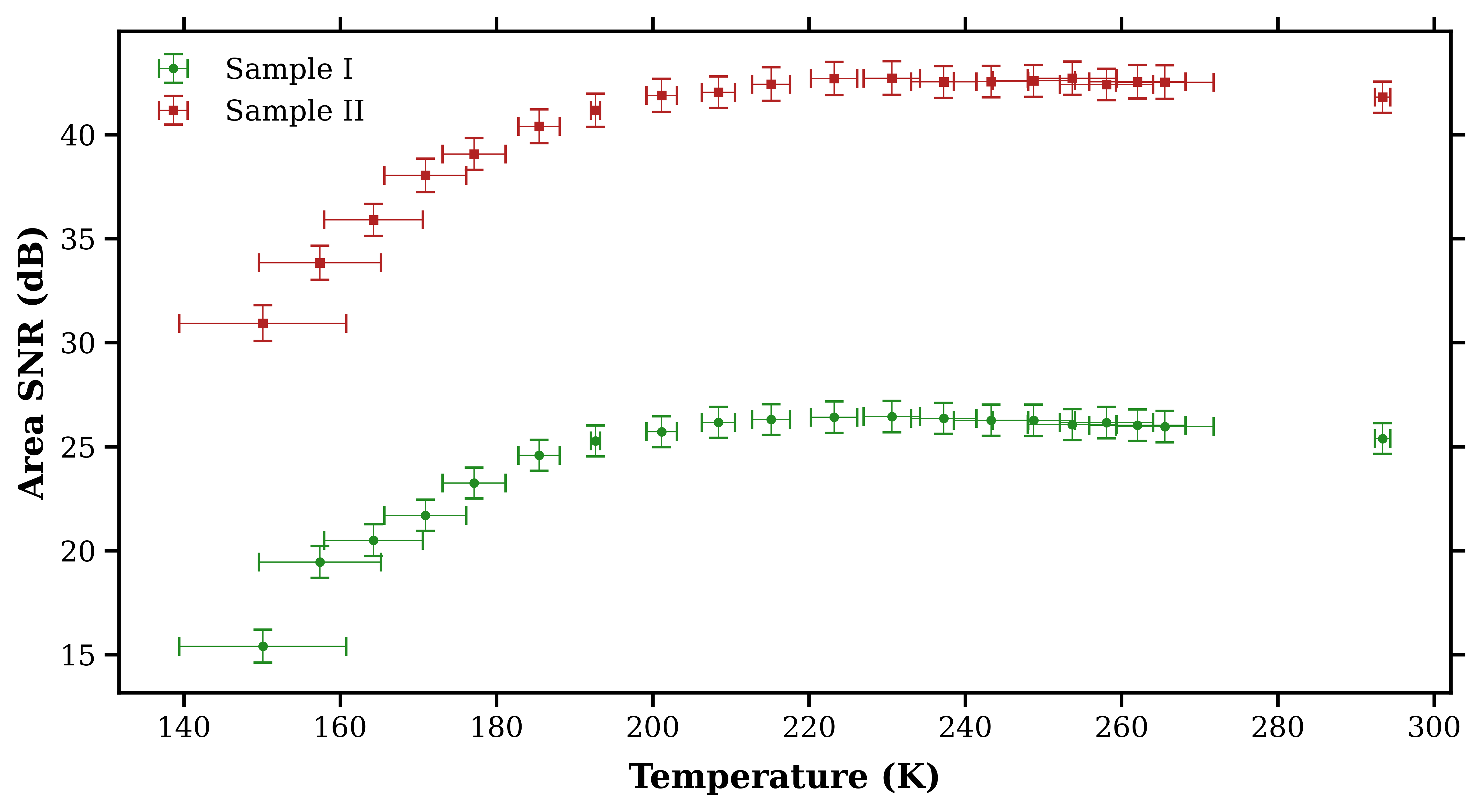}
\caption{Comparison of the area (or power) SNR for Samples I and II as a function of temperature.}
\label{fig:cryo_SNR_comparison}
\end{figure}

The Signal-to-Noise Ratio (SNR --- shown in Figure \ref{fig:cryo_SNR_comparison}) is determined by dividing the area under the first peak by that under the same frequency region of the experimentally determined background spectrum, as described in Equation \ref{eq:SNR}. 

\noindent
\begin{equation}
SNR_{area} = \frac{\int_R PSD_{sig}(f)\, df
}{\int_R PSD_{bkg}(f)\, df} = \frac{A_{sig}}{A_{bkg}}
\label{eq:SNR}
\end{equation}

where $R$ represents the region around the fundamental frequency (shaded in figure \ref{fig:cryo_data_frequency_domain}). This region was defined by taking a tolerance of 0.02~Hz around the first peak (to well-contain the peak region of the measured signal in the frequency domain).

To determine the temperature of the sample during these low-temperature measurement runs, the temperature calibration curve measured with the RTD in front of the Cu block is utilized. For each point in time, the temperature is considered to come from a normal distribution, and, for each run, these normal distributions are averaged across the period of time used for that signal measurement, weighted by the Hann window defined in Appendix 1 Equation \ref{eq:defining w}. The mean and standard deviation of each resulting distribution are used to plot the temperature axis and the temperature uncertainties in Figure \ref{fig:cryo_SNR_comparison}. The error in the  SNR is obtained by considering the error in the calculation of the area under the first peak in the power spectrum as shown in Figure \ref{fig:cryo_data_frequency_domain} and the fluctuations in the background power spectrum during each measurement of the detector response to the UV-LED pulsing. The calculation of the error on the SNR is described in Appendix 1. 

Our measurement shows an enhancement beyond the measurement errors in the area SNR plot (Figure \ref{fig:cryo_SNR_comparison}) for Sample II in comparison to Sample I. The enhancement is distinctively present even in the lowest temperature achievable with the in-house cryogenic test-stand. This suggests a considerably well-behaved performance of the ZnO-based UV-detectors in low-temperature environments. Further analysis of the device architecture and different ZnO-doped samples can be explored to investigate the SNR enhancement of these detectors.

\subsection{Residual Strain using Grazing-Incidence X-Ray Diffraction}

To characterize the stress induced in the interfacial layer between ZnO and Si\textsubscript{3}N\textsubscript{4}, a Grazing Incidence XRD (GIXRD) was performed using a Bruker D8 Advance~\cite{BrukerD8Advance}. 

\begin{figure}[htbp]
\centering

\begin{subfigure}[t]{\textwidth}
    \includegraphics[width=\textwidth]{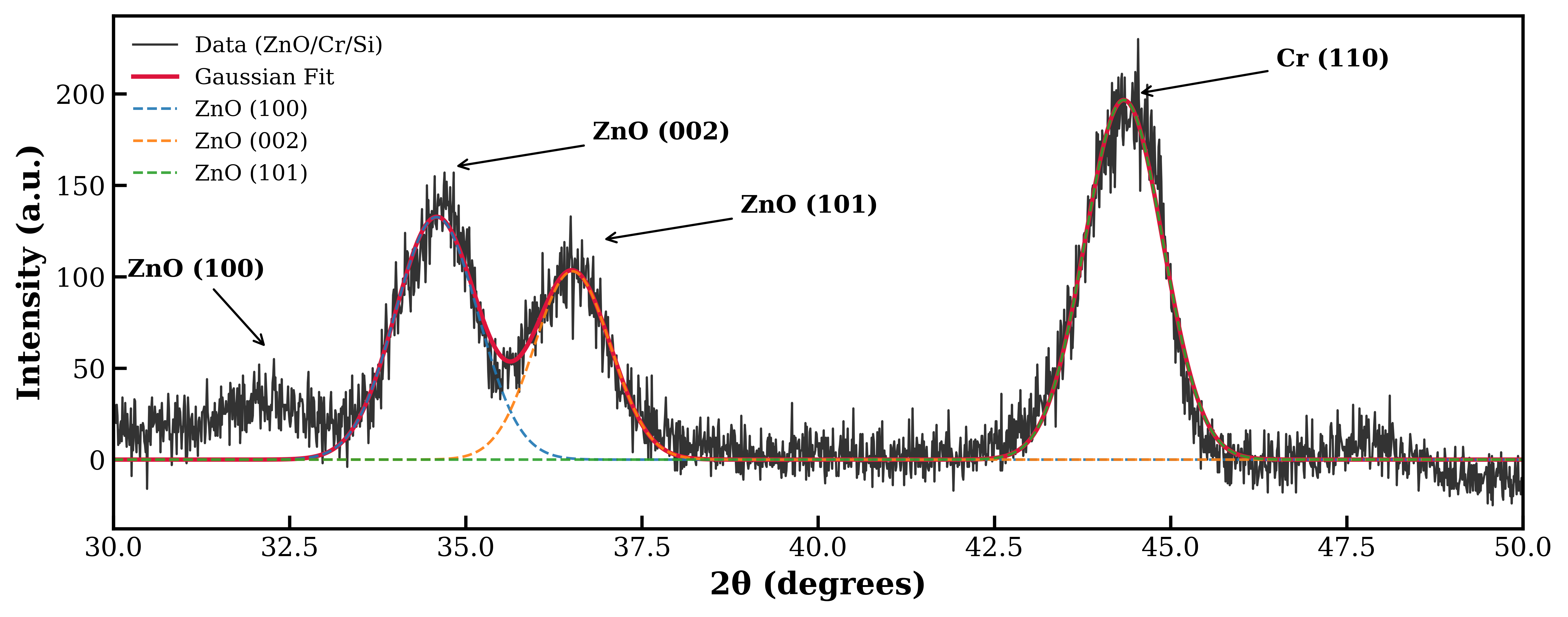}
    \caption{GIXRD spectrum of the ZnO/Cr/Si stack.}
\end{subfigure}

\vspace{0.5em} 

\begin{subfigure}[t]{\textwidth}
    \includegraphics[width=\textwidth]{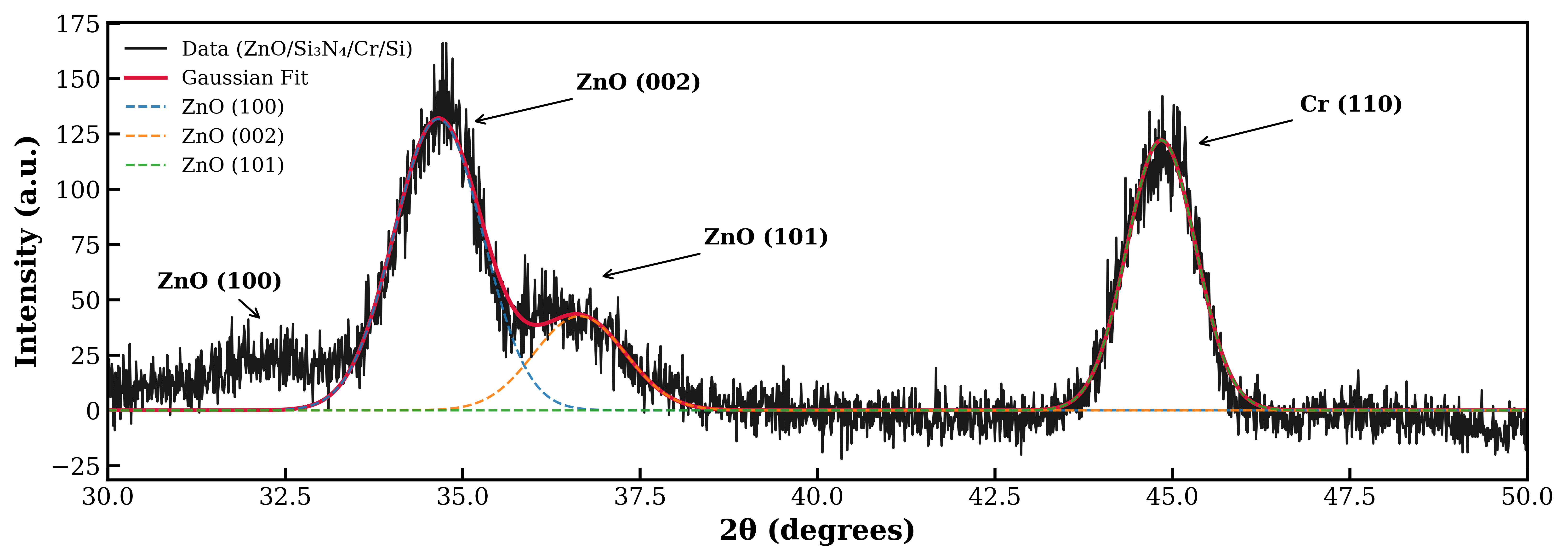}
    \caption{GIXRD spectrum of the ZnO/Si\textsubscript{3}N\textsubscript{4}/Cr/Si stack.}
\end{subfigure}

\caption{The GIXRD spectra for a bare ZnO and a ZnO–Si\textsubscript{3}N\textsubscript{4} heterostructure stack are shown, (a) and (b) respectively. The tube was set at a shallow angle of $0.1 ^\circ$ to collect the spectra. The ZnO (002), (101), and Cr (110) peaks have been fit with Gaussian distributions. The peak center and FWHM of the ZnO (002) peak were used to determine the induced residual stress due to lattice mismatch. The instrumental broadening was initially determined to be $0.27 ^\circ$ Full-Width at Half Maximum (FWHM) using an intrinsic Si sample.}
\label{fig:GIXRDplots}
\end{figure}

\FloatBarrier

In Figure~\ref{fig:GIXRDplots} the GIXRD spectra for the Sample I (top) and Sample II (bottom) were obtained using an incident X-Ray (1.5406 Å) angle of 0.1$^\circ$. We observe that the primary peak for the ZnO is seen to be along the (002) orientation. This confirms the major c-axis growth of the ZnO layer. Gaussian distributions were fit to the observed primary peaks to obtain the Full-Width-At-Half-Maximum (FWHM) and the centers of these peaks. The instrumental broadening of the XRD apparatus was first determined by collecting the spectra for a clean intrinsic Si wafer. This instrumental broadening was found to be $0.27 ^\circ$ FWHM. Additionally, the crystalline size of the ZnO crystals were determined using a Hitachi S-4800 FESEM as shown in Figure~\ref{fig:csSEMfigures}  and listed in Table~\ref{tab:gixrd}. Hence, considering the instrumental resolution and the crystalline size, the residual stress in the material - attributed to both localized microstructural distortions and lattice mismatch was determined using the following Equations~\ref{eq:total_broadening} – \ref{eq:lattice_strain}: 

\begin{equation}
\beta_t^2 = \beta_D^2 + \beta_S^2 + \beta_0^2
\label{eq:total_broadening}
\end{equation}

\begin{align}
\beta_D &= \frac{K \lambda}{D \cos \Theta} && \text{(Size broadening)} \\
\beta_S &= 4 \epsilon_{\text{micro}} \tan \Theta && \text{(Microstrain broadening)} \\
\beta_0 &= \text{Instrumental broadening (constant)}
\end{align}

where $K$ is the shape factor (0.9); $\lambda$ refers to the wavelength of the incident X-ray; $D$ is the crystalline size; $\Theta$ refers to the Bragg's angle.

Solving for microstrain \( \epsilon_{\text{micro}} \):

\begin{equation}
\epsilon_{\text{micro}} = \frac{1}{4 \tan \Theta} \sqrt{\beta_t^2 - \beta_D^2 - \beta_0^2}
\label{eq:microstrain}
\end{equation}

The lattice strain \( \epsilon_{\text{lattice}} \), obtained from the shift in peak position, is calculated using the difference in the measured and reference interplanar spacing \( d \) (or lattice parameter \( c \)):

\begin{equation}
\epsilon_{\text{lattice}} = \frac{d_{\text{measured}} - d_{\text{reference}}}{d_{\text{reference}}}
\label{eq:lattice_strain_theory}
\end{equation}

For example, in the case of a hexagonal or layered material (e.g., ZnO), the out-of-plane lattice parameter \( c \) can be used:

\begin{equation}
c_{\text{measured}} = \frac{n \lambda}{\sin \Theta}, \quad \epsilon_{\text{lattice}} = \frac{c_{\text{measured}} - c_0}{c_0}
\label{eq:lattice_strain}
\end{equation}

The theoretical value of $c_0$ for ZnO, used in this study was 5.21, as determined from the Miller indices~\cite{ennaceri2019influence}. Both Sample I and Sample II show a compressive lattice stress of approximately 1.08 GPa and 1.53 GPa, respectively. Additonally, the microstrain of the two samples are 0.12 and 0.11 respectively. Hence, we observe that due to the introduction of the Si\textsubscript{3}N\textsubscript{4} layer, the residual compressive stress increases by $42.7 \%$. Table~\ref{tab:gixrd} summarizes the material characteristic determined from the GIXRD measurements of the samples. 

\begin{table}[htbp]
\centering
\resizebox{\textwidth}{!}{%
\begin{tabular}{l|c|c|c|c|c|c|c|c}
\hline
Sample & Crystallize Size (nm) & FWHM & Peak Center & d-spacing & C\textsubscript{0} (theoretical) & C (measured) & $\varepsilon_{zz, lattice}$ (\%) & Residual Stress (GPa) \\
\hline
Sample I  & 50  & 1.56 & 34.59 & 2.59 & 5.21 & 5.18 & -0.46 (compressive) & 1.08 \\
Sample II &  50 & 1.47 & 34.66 & 2.58 & 5.21 & 5.17 & -0.66 (compressive) & 1.53 \\
\hline
\end{tabular}%
}
\smallskip
\caption{Material characteristics obtained from Sample I and Sample II using a 1.5406\,\AA{} X-ray with an incident angle of 0.1 degree. The crystalline size was determined from FeSEM measurements.
\label{tab:gixrd}}
\end{table}

\section{Conclusions}

In this paper, we demonstrate the fabrication and characterization of residual stress enhancement achieved by incorporating Si\textsubscript{3}N\textsubscript{4}  pillars, for their operation as a UV-detector for cryogenic conditions. The devices employ ZnO thin films as the active semiconductor layer within a MSM architecture. The thin-film devices were fabricated using plasma sputtering techniques integrated with shadow-masked depositions for achieving the pillar and metal electrode geometries. The UV-induced signal performance of these fabricated devices were characterized in vacuum and across different temperatures, in a custom-built test-stand. A 365 nm UV-LED light source was used to illuminate the active region of the devices mounted in the test-stand. We observe an enhanced UV-induced voltage signal for Sample II compared to Sample I, at room temperature, and at varying light intensities of the LED. Additionally, a pulsed UV-assisted study paired with dedicated signal processing techniques were implemented to characterize the low-absolute signal intensities observed in cryogenic conditions. The obtained SNR values of the samples suggest a distinguishable enhancement observed in Sample II in comparison to Sample I. 

The residual stress induced in these devices were quantified using GiXRD measurements. The analysis of these measured diffraction peaks coupled with FeSEM measurements of the particle sizes, show a compressive stress of 1.58 GPa for Sample II in comparison to 1.03 GPa for Sample I. This is expected to produce a higher piezo potential in the ZnO layer of Sample II. This enhanced piezo-induced electric field assists in transporting the photo-generated carriers created due to the absorption of UV light, thereby leading to an enhanced UV detection sensitivity for the case of Sample II in comparison to Sample I. Our studies provide indirect, but repeatable evidence for stress induced piezo enhancement of the UV induced photo signal characteristics like faster rise time and higher signal amplitude. Future experiments are planned that provide a direct correlation between the increase in piezo potential to the increase in UV induced signal. These studies show the possible application of such ZnO-based UV detectors in extreme environments. The ease of fabrication, cost-effectiveness, and tunable electromechanical properties of these devices present opportunities for further optimization, including alternative device geometries and controlled doping strategies to improve carrier transport and sensitivity.

\acknowledgments

This work is supported by Department of Energy and the Office of Science under award number: DE-0000253485. The work is also supported by the Nanotechnology Research Center, Shimadzu Institute, UTA. The authors also acknowledge the support from the Characterization Center for Materials and Biology at UTA, where the FeSEM and GiXRD measurements were performed.

\begin{appendices}
\section{Additional information}
\subsection{Data Analysis}
\label{appendix:Data Analysis}
\setcounter{equation}{0} 
\renewcommand{\theequation}{A.\arabic{equation}} 

\setcounter{figure}{0}
\renewcommand{\thefigure}{A. \arabic{figure}}

 \begin{figure}[H]
     \includegraphics[width=\textwidth]{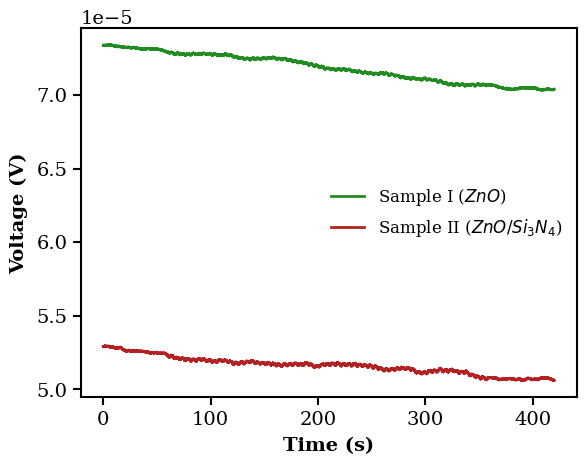}
 \caption{Raw data measured as described in Section \ref{sec:lowTempMeasurement} using the nanovoltmeter at 164 ± 6 K.}
 \label{fig:raw_data}
 \end{figure}

\begin{figure}[H]
 \centering
     \includegraphics[width=\textwidth]{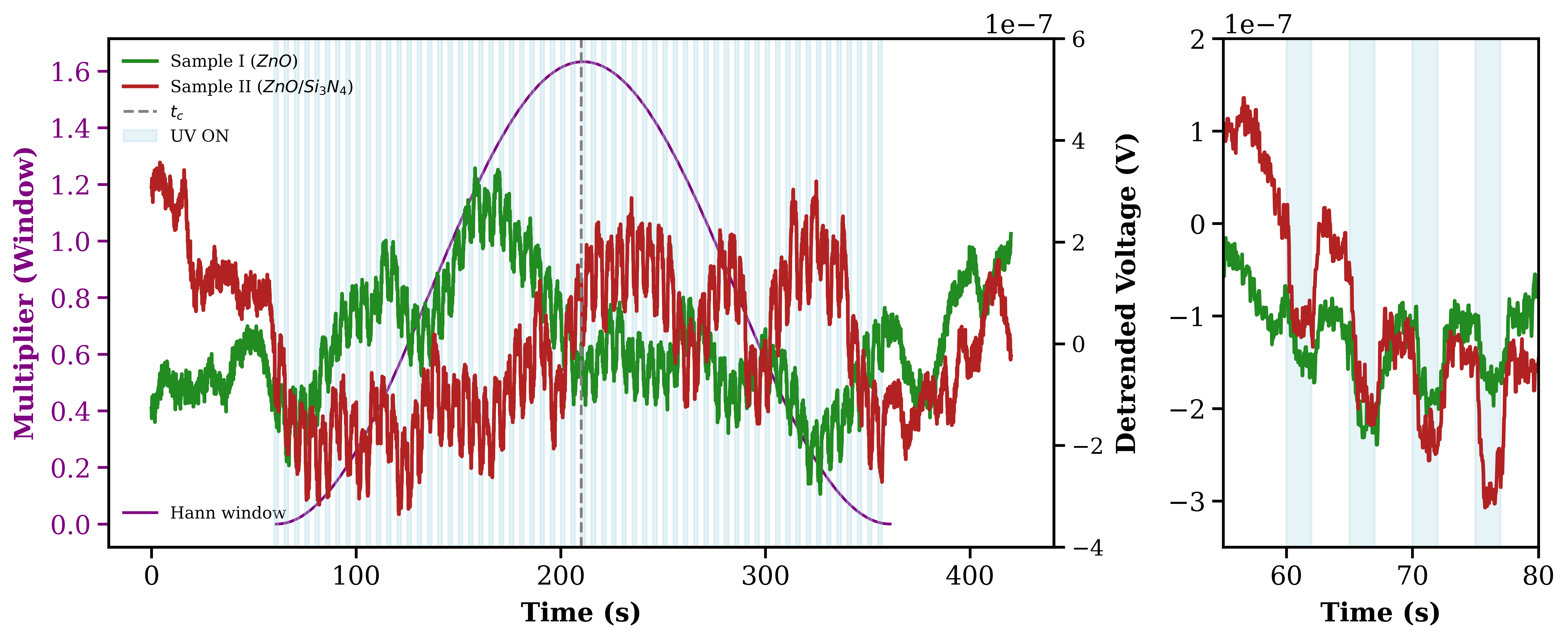}

 \caption{The detrended time domain voltage signals D(t) as given in Equation \ref{eq:detrended signal} received from Samples I and II at a temperature of 164~$\pm$~6 K. The shaded region shows the time window ($L$) during which the UV-LED was pulsed. $t_{c}$ corresponds to the temporal midpoint of this pulsing interval. The Hann window function the signal was multiplied by is shown in purple. It is scaled to have its root mean square calculated over the time period $L$ unity so as to leave the average power unchanged during this period. At right are shown the responses to the first four UV pulses so that the UV-light-induced responses can be seen more clearly.}
 \label{fig:cryo_data_time_domain}
 \end{figure}

Figure \ref{fig:raw_data} shows the voltage as a function of time (V(t)) measured using the nanovoltmeter. The first step of the analysis is to remove a linear trend from the data. The liner trend is caused by the slowly varying temperature during the course of temperature dependent measurement. The detrending also corrects for any variation in the baseline voltage for the two samples.  We detrend by subtracting a straight line obtained through a linear fit to each signal ($y(t)$ in Equation \ref{eq:detrended signal}). Figure \ref{fig:cryo_data_time_domain} shows one such detrended time domain signal for each sample.

\noindent
\begin{equation}
D(t) = V(t) - y(t)
\label{eq:detrended signal}
\end{equation}

After obtaining the detrended signal, a Hann (or Hanning) window \cite{Blackman1958Windowing} is applied to the signal region during which
the UV-LED was pulsed, followed by a high pass filter to remove the noise below 0.15~Hz, as described in Equation \ref{eq:processed signal}. Here $\mathcal F$ represents a Fourier transform and H represents the Heaviside step function. Equation \ref{eq:processed signal} written in expanded form using the definition of the Discrete Fourier Transform (DFT) is given in Equation \ref{eq:processed signal discrete} 

\noindent
 \begin{equation}
 S_{WH}(t) = \mathcal{F}^{-1}\{H(f-0.15\text{~Hz})\mathcal{F}\{w(t)D(t)\}(f)\}(t)
 \label{eq:processed signal}
 \end{equation}
 \noindent 
 
\noindent
\begin{equation}
S_{WH}(t_n) =\frac{1}{N}\sum_{k=0}^{N-1}e^{i 2 \pi k n / N}H(f_k-0.15\text{~Hz})\sum_{m=0}^{N-1}e^{-i 2 \pi k m / N}w(t_m)D(t_m)
\label{eq:processed signal discrete}
\end{equation}
\noindent 

where,

\noindent
\begin{equation}
f_k =\frac{k}{L}.
\label{eq:defining f_k}
\end{equation}
\noindent

Here, $w(t)$ is the Hann window function which is applied to the signal over the total time window ($L$) where the UV was pulsed and $N$ is the total data length of $S_{WH}$.This windowing is implemented to control spectral leakage and reduce scalloping loss (essentially to mitigate the amplitude loss of the signal when its frequency lies between the DFT bins), which is otherwise observed in the Fourier transformed signals in some of the low-temperature runs. The Hann window function was normalized to have its root mean square calculated over $L$ to be unity to keep the average power in the measured signal unchanged due to the application of the  windowing function.  

\begin{equation}
w(t) =
\begin{cases}
  \sqrt{\frac{8}{3}}\cos^{2}(\frac{\pi}{L}(t-t_{c})), & \text{if } |t-t_{c}|<L/2 \\
  0,  & \text{if } |t-t_{c}| \geq L/2
\end{cases}
\label{eq:defining w}
\end{equation}

$H(x)$ is the Heaviside function which represents the action of the high pass filtering we performed in the frequency domain.

\begin{equation}
H(x) =
\begin{cases}
  0, & \text{if } x<0 \\
  1,  & \text{if } x \geq 0
\end{cases}
\label{eq:defining H}
\end{equation}

\begin{figure}[H]
 \centering
     \includegraphics[width=\textwidth]{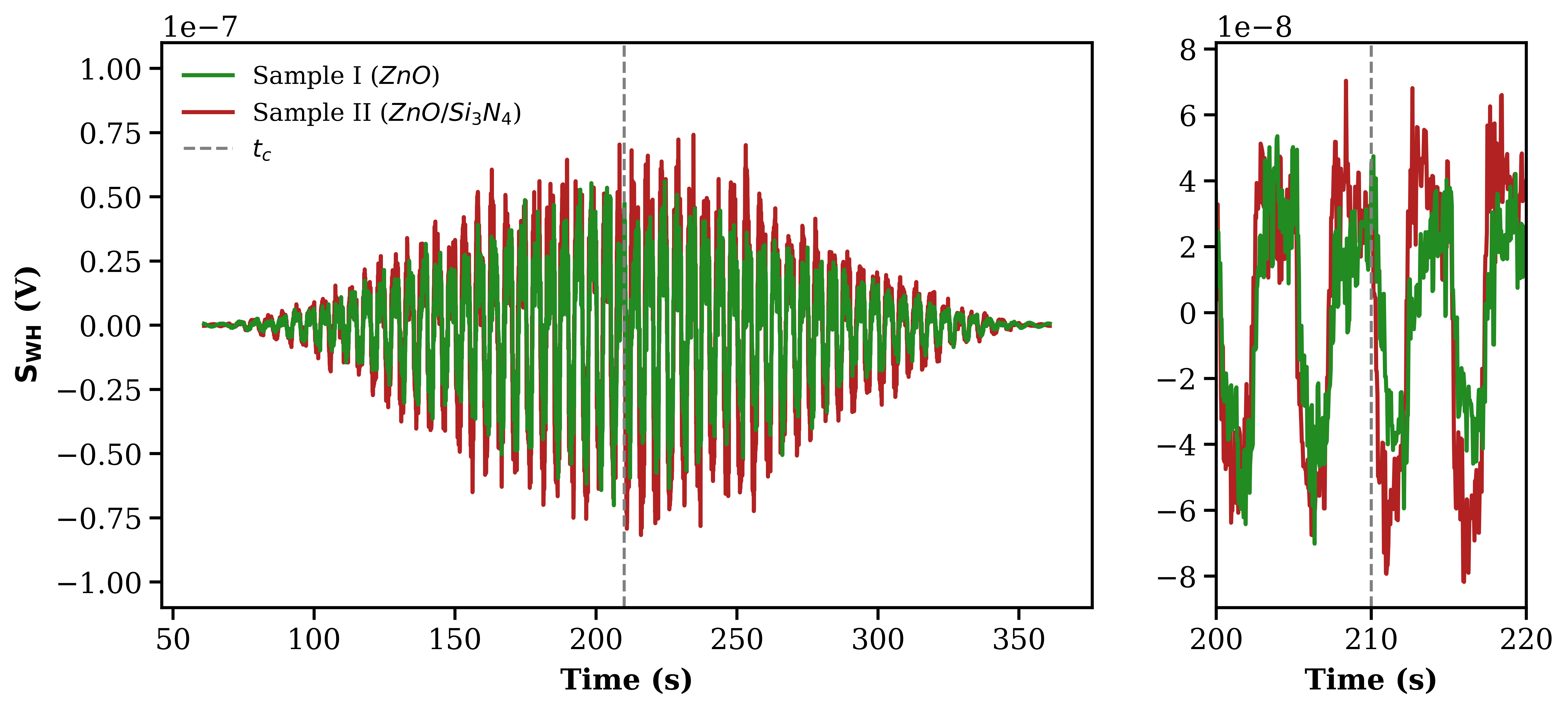}
 \caption{The windowed high-passed signal ($S_{WH}$ from Equation \ref{eq:processed signal discrete}) for both samples is depicted. The high-pass filter removes slow oscillations. The Hann window is applied only over the time period when the UV-LED was pulsed and it results in a signal which looks as if it is amplitude modulated by the Hann Window envelope. The spectral broadening as a result of Hann windowing is predictable and is dependent on the shape of the window chosen. On the right, the time axis is zoomed in to show the UV response of the samples. Here the response difference between the samples are clearly more visible.}
 \label{fig:S_WH}
 \end{figure}

 The Hann-windowed and high-pass-filtered signal for measurement at 164~$\pm$~6 K  is shown in Figure \ref{fig:S_WH}. 

\begin{figure}[htb!]
 \centering
     \includegraphics[width=\textwidth]{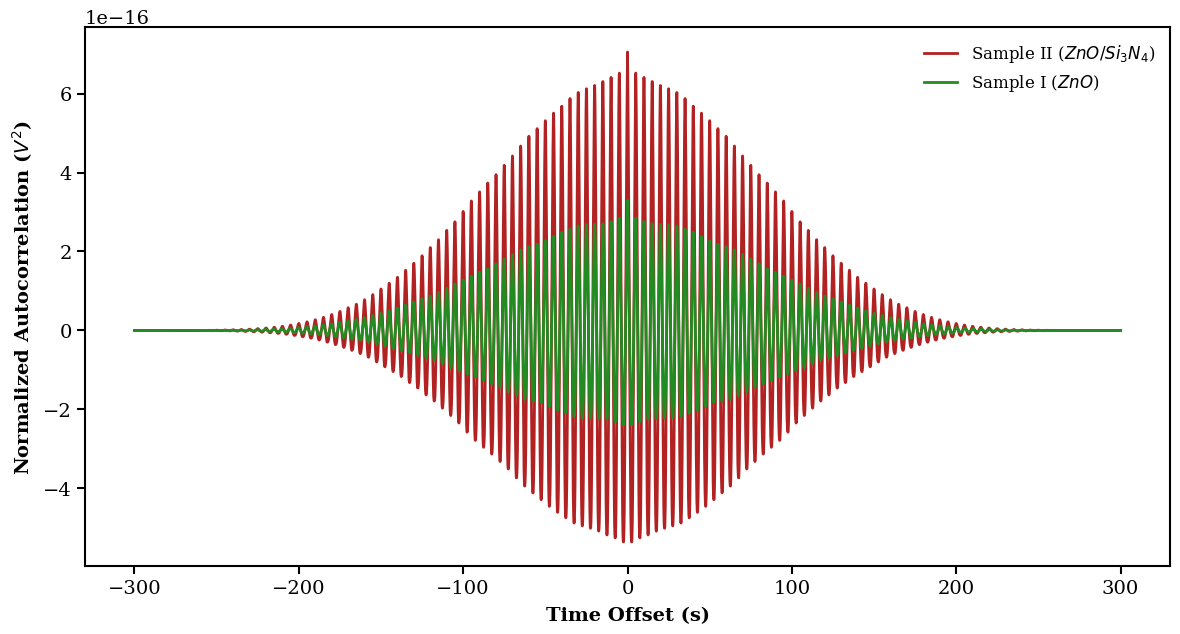}
 \caption{Normalized Autocorrelation $S_{AC}(t)$ of the detrended, Hann windowed, high pass filtered time domain signals $S_{WH}(t)$ obtained from Sample I and II at 164~$\pm$~6 K. The autocorrelation enhances the difference in response of the two devices and helps us to calculate the PSD}
 \label{fig:autocorrelation}
 \end{figure}

The Wiener-Khinchin (WK) theorem states that the power spectral density (PSD) of any temporally independent
and identically distributed signal is the Fourier transform of its autocorrelation function \cite {Wiener_1930, lu2009wienerkhinchintheoremnonwidesense}. We apply the WK theorem to obtain the power generated by our device when it is exposed to repeated UV pulses. For this, we use the detrended Hann-windowed high-pass-filtered signal, $S_{WH}(t)$, to calculate the power output of our device per unit frequency multiplied by the impedance across the chromium terminals during measurements. That is, for our data, PSD is expressed as 

\begin{equation}
PSD(f) = \mathcal{F}\{S_{WH}(t) \star S_{WH}(t)\}
= \int_{-\infty}^{\infty} {e^{-i2\pi f \tau} \int_{-\infty}^{\infty}{S_{WH}(t) S_{WH}(t+\tau)\,dt}\,d\tau}
\label{eq:PSD signal}
\end{equation}

where $S_{WH}(t) \star S_{WH}(t)$ represents the normalized autocorrelation of $S_{WH}(t)$. The normalized autocorrelation ($S_{AC}$) for our discrete signal was calculated using Equation \ref{eq:defining S_AC} and shown in Figure \ref{fig:autocorrelation}.

\begin{equation}
S_{AC}(\tau_n) = \frac{1}{N}\sum_{m=0}^{N-1}{S_{WH}(t_m)S_{WH}^*(t_{m-n})}
\label{eq:defining S_AC}
\end{equation}  

Here we assume the time offset $\tau_n=t_m-t_{m-n}$ is constant for all m due to the approximately even spacing of the voltage samples in time. Indices out of range are considered to make $S_{WH}$ equal zero. The fact that the complex conjugate of $S_{WH}(t_{m-n})$ is taken is not important because it is always a real number in our case.

PSD \cite{Paschotta_2005_power_spectral_density} can then be calculated by taking the DFT of $S_{AC}$ as in Equation \ref{eq:PSD calculation} and is shown in Figure \ref{fig:cryo_data_frequency_domain}

\begin{equation}
PSD(f_k) = \frac{2}{(2N-1)\Delta f}\left\lvert\sum_{n=1-N}^{N-1}{e^{-i2\pi k(n+N-1)/(2N-1)}S_{AC}(\tau_n)}\right\rvert
\label{eq:PSD calculation}
\end{equation}

\begin{figure}[H]
 \centering
 \includegraphics[width=\textwidth]{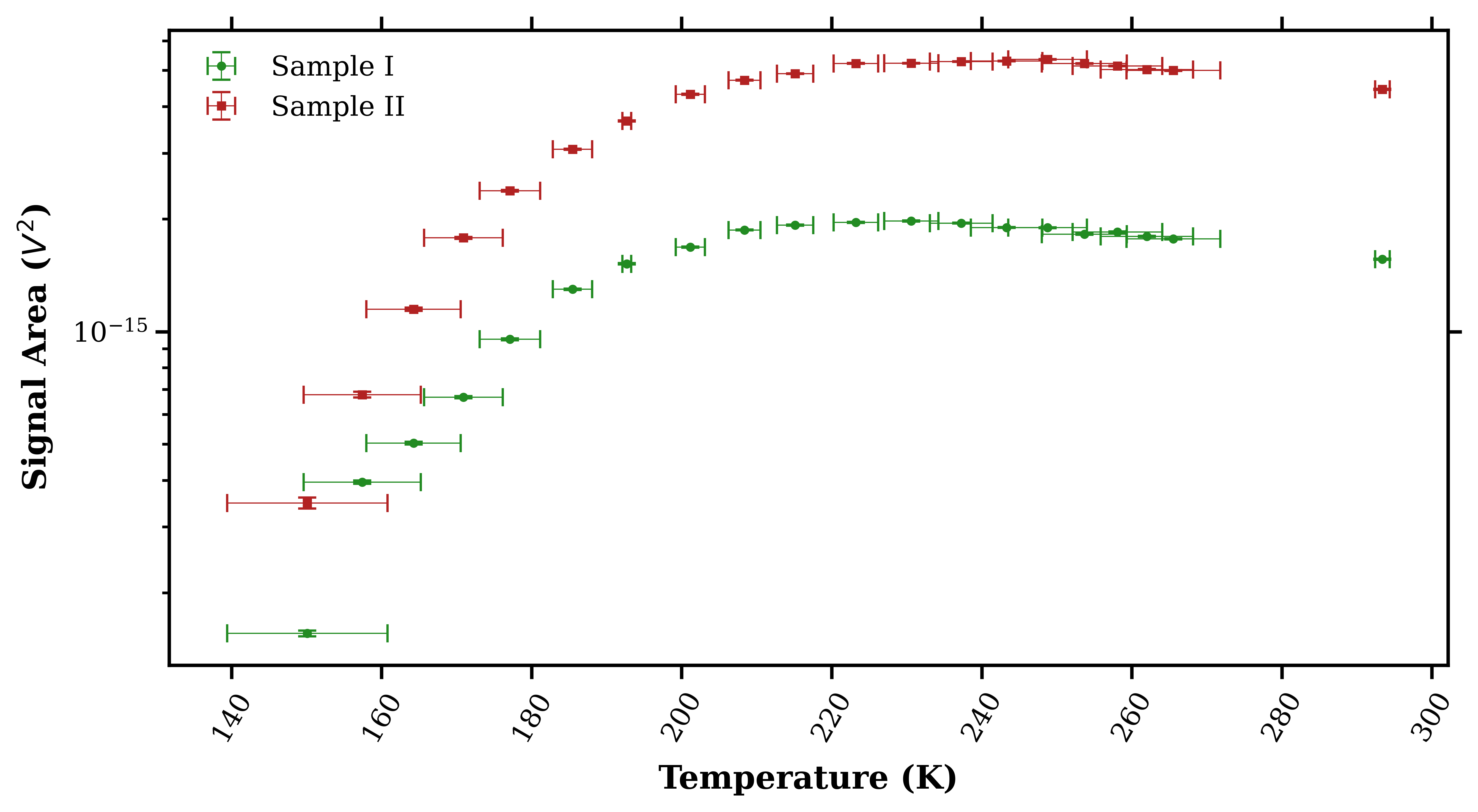}
 \caption{Comparison of the total area under the fundamental frequency $(A_{sig})$ for Samples I and II as a function of temperature. The errors on the area are overestimates calculated using Equation \ref{eq:total S_WH power error}.}
 \label{fig:area_sig_comparison}
 \end{figure}

Please note that $\Delta f$ in Equation \ref{eq:PSD calculation} is the reciprocal of $(2N-1)\Delta t$, where $\Delta t$ is the mean sampling interval in our measurement. From the PSD, we calculate the SNR for each device considering the area under the fundamental frequency (the first peak) as given by Equation \ref{eq:SNR}. To calculate the area under the frequency range ($R$) of interest for both the UV generated signal and for the background we used the following Riemann sum:

\begin{equation}
A = \sum_{\substack{f_k \in R \\ k \in \mathbb{Z}}}{PSD(f_k)\Delta f}
\label{eq:Area calculation}
\end{equation}.

where $\mathbb{Z}$ refers to the set of integers. The Area of the first peak for Sample I and II are shown in Figure \ref{fig:area_sig_comparison}.

\begin{figure}[H]
 \centering
 \includegraphics[width=\textwidth]{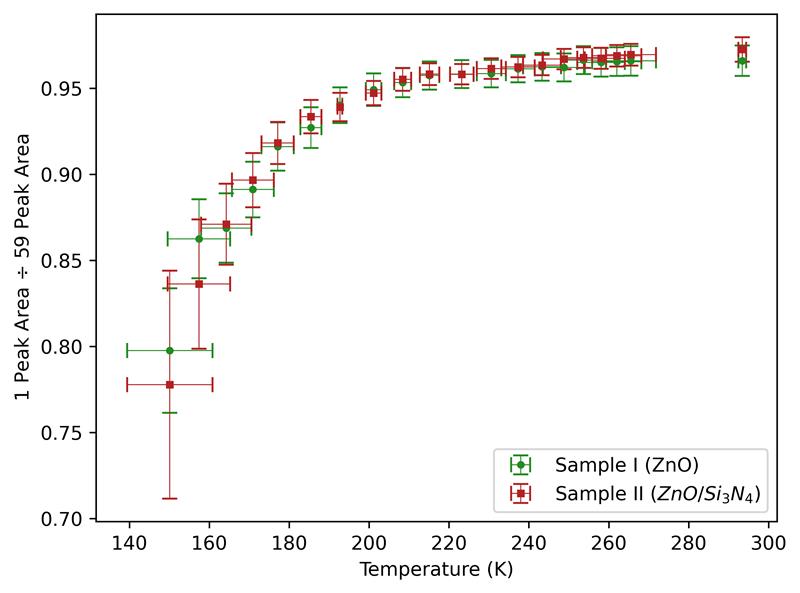}
 \caption{Ratio of the area under the fundamental frequency $(A_{sig})$ to the total area under all harmonics accessible in the PSD for Samples I and II as a function of temperature. Our analysis shows that majority of the power is within the first peak, and hence only that is considered for the calculation of the SNR. }
 \label{fig:first_peak_area_sig_comparison}
 \end{figure}

We considered only the area under the first peak since majority of the power is within the fundamental frequency as shown in Figure \ref{fig:first_peak_area_sig_comparison}. Here we have compared the area under first peak to the sum of the areas under the first peak and all the higher harmonics below the $\sim$12 Hz Nyquist frequency (59 harmonics total, most of which do not have a peak visible above background). The calculated SNR is shown in Figure \ref{fig:cryo_SNR_comparison}.

\subsection{Calculation of Error on the SNR obtained from PSD}
\label{appendix:Error Propagation}

The calculation of the error in the SNR determined using Equation \ref{eq:SNR} must consider the error on $A_{sig}$ and $A_{bkg}$. $A_{sig}$, determined from the area under the peak corresponding to the fundamental frequency in the PSD (shown in Figure \ref{fig:cryo_data_frequency_domain} and using Equation \ref{eq:Area calculation}),  has a unit of $volt^2$, which is power ($P$) times the impedance ($Z$) of the respective device. The error in $A_{sig}$ should therefore consider the fluctuations on the measurement of the photocurrent generated by the device using the nanovoltmeter. These voltage fluctuations seem significant when comparing the output of the two devices under a single "on" pulse (see the zoomed in time-domain signal shown in the right panel of Figure \ref{fig:S_WH}). However, the average $P \times Z$ calculated over several "on" pulses found using the area under the PSD shows significant enhancement beyond the voltage fluctuations. Ideally, to determine the error on $A_{sig}$, these voltage fluctuations should be propagated through the autocorrelation operation, the Fourier transform operation, and the integration over the frequency range that we applied to determine $A_{sig}$. One advantage of the frequency domain analysis is the ability to select the signal over a specific frequency range ($R$) allowing us to remove higher frequency noise which becomes significant if the signal is very small compared to the voltage fluctuations. However, we can also calculate the average total $P \times Z$ (over all frequencies from 0.15 Hz to $\sim$12 Hz) directly from the Hann windowed, high pass filtered time-domain signal $S_{WH}$ using the equation:

\noindent
\begin{equation}
P\times Z =\frac{1}{N}\sum_{m=0}^{N-1}{(S_{WH}(t_m))^2}
\label{eq:total_S_WH power}
\end{equation}
\noindent 

Applying the standard error propagation formulae \cite{Ku_1966}, the uncertainty on this $P \times Z$ is:

\noindent
\begin{equation}
\sigma_{PZ} =\frac{1}{N}\sqrt{\sum_{m=0}^{N-1}{[2S_{WH}(t_m)]^2var[S_{WH}(t_m)]}}
\label{eq:total S_WH power error}
\end{equation}.
\noindent

Here, the covariances between the $S_{WH}$ voltages at different times have been neglected because, visually, the rough shape of the signals imply these covariances are small, and, theoretically, the shot noise on adjacent voltage measurements should not be strongly correlated. The variances of the individual $S_{WH}$ values were assumed to depend only on the phase in the 5 s periodic waveform and on the multiplication by the Hann window function. For a particular $t_m$ in a particular run, the $var[S_{WH}(t_m)]$  was calculated in the following way:

\noindent
\begin{equation}
var[S_{WH}(t_m)]=w^2(t_m)var\left[\frac{S_{WH}(t_m)}{w(t_m)}\right]
\label{eq:total S_WH variances}
\end{equation}.
\noindent

where $var\left[\frac{S_{WH}(t_m)}{w(t_m)}\right]$ is calculated by comparing the values of $\frac{S_{WH}(t_{m'})}{w(t_{{m'}})}$ for most of the $m'$ in the run such that $t_{m'} \approx t_m ~mod~5 s$. (The edges of the window were excluded due to numerical instability in dividing by the window function.) This was done because $\frac{S_{WH}(t_m)}{w(t_m)}$ is approximately a periodic function plus noise. The resulting standard deviations on individual $\frac{S_{WH}(t_m)}{w(t_m)}$ values were all similar in magnitude, mostly staying between $\sim$5e-9 V and $\sim$1e-8 V.  

\begin{figure}[H]
 \centering
 \includegraphics[width=\textwidth]{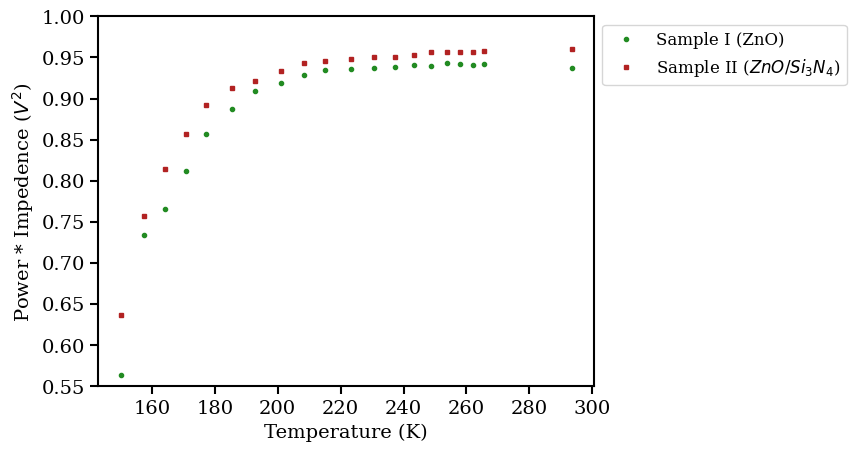}
 \caption{Ratio of the area under the fundamental frequency $(A_{sig})$ to the total power of the $S_{WH}$ signal.}
 \label{fig:first_peak_S_WH_area_comparison}
 \end{figure}

$P \times Z$ calculated from the time-domain signal $S_{WH}$ includes the photoinduced power as well as the power in the noise, whereas that calculated from the single peak in PSD ($A_{sig}$) has significantly less contribution from higher frequency noise. For this reason, the $P\times Z$ calculated using Equation \ref{eq:total_S_WH power} will be greater than the $A_{sig}$ calculated using Equation \ref{eq:Area calculation}, and they are equal only when $R = [0,f_{max}]$ where $f_{max}=\frac{1}{2\Delta t}\approx12~\text{Hz}$ is the Nyquist frequency. The ratio of the area under the fundamental frequency ($A_{sig}$) to the total $P \times Z$ calculated using the time domain signals $S_{WH}$ generated by Samples I and II is shown in Figure \ref{fig:first_peak_S_WH_area_comparison}. 

Total $P \times Z$ can also be calculated from the PSD using Equation \ref{eq:PZ from PSD}, which yields the exact same answers as Equation \ref{eq:total_S_WH power} up to rounding errors.

\noindent
\begin{equation}
P \times Z ~~~= \sum_{\substack{f_k \in [0,f_{max}] \\ k \in \mathbb{Z}}}{PSD(f_k)\Delta f} = \sum_{k=0}^{N-1}{PSD(f_k)\Delta f}
\label{eq:PZ from PSD}
\end{equation}

Thus, considering Equations \ref{eq:Area calculation} and \ref{eq:PZ from PSD} and the standard error propagation formulae \cite{Ku_1966}, the errors on $A_{sig}$ and $P \times Z$ can be theoretically written like so:

\noindent
\begin{equation}
\sigma_{A_{sig}}=\Delta f \sqrt{\sum_{\substack{f_k,f_l \in R \\ k, l \in \mathbb{Z}}} cov[PSD(f_k),PSD(f_l)]}
\label{eq:sigma_A_sig}
\end{equation}

\noindent
\begin{equation}
\sigma_{PZ}=\Delta f \sqrt{\sum_{k=0}^{N-1} \sum_{l=0}^{N-1} cov[PSD(f_k),PSD(f_l)]}
\label{eq:sigma_PZ}
\end{equation}

The sum in Equation \ref{eq:sigma_PZ} includes all the terms that are summed over in Equation \ref{eq:sigma_A_sig}, plus many more. Thus, unless the covariances between the PSDs in different frequency bins are large and predominantly negative (which would mean that more power in one frequency range implies less power in other frequency ranges), $\sigma_{PZ} \geq \sigma_{A_{sig}}$. We expect predominantly positive covariances between nearby frequency bins due to the smoothing effects of spectral broadening on the PSD and see no reason to expect any negative covariances between distant frequency bins to outweigh this effect. Therefore, we have used $\sigma_{PZ}$ (the error on $P \times Z$) as an upper bound on $\sigma_{A_{sig}}$ (the error of $A_{sig}$) to calculate the error on the SNR.  

To estimate the error on $A_{bkg}$, we used the equation: 

\noindent
\begin{equation}
\sigma_{PSD}^2(f) = \left(\frac{1}{\#-2}\sum_{k=0}^{\#}\left[PSD_{bkg}(f_k)-PSD_{fit}(f_k)\right]^2\right)
\label{eq:stdPSD}
\end{equation}

where $\#$ is the number of frequency bins that lie in $R$, $PSD_{bkg}$ is the PSD of the experimental background spectrum, and $PSD_{fit}$ is a "fit" to $PSD_{bkg}$. This "fit" is a least squares fit of a linear function to the logarithm of $PSD_{bkg}$ as a function of the logarithm of frequency (and so has two degrees of freedom). This error was then propagated through the Riemann sum given in Equation \ref{eq:Area calculation} to obtain $\sigma_{A_{bkg}}$ as shown in Equation \ref{eq:stdA}.

\noindent
\begin{equation}
\sigma_{A_{bkg}}^2 = \frac{{0.4~Hz}~^2}{\#}\sigma_{PSD}^2
\label{eq:stdA}
\end{equation}

$0.4 \text{~Hz}$ is the width of frequency range ($R$) under the first peak over which area summation was done. 

The error on the SNR are then calculated from this using standard error propagation formulae \cite{Ku_1966}, assuming the background in each run to be independent of the other runs, as shown in Equation \ref{eq:stdSNR}.

\noindent
\begin{equation}
\sigma_{SNR_{area}} = \frac{A_{sig}}{A_{bkg}} \sqrt{\frac{\sigma_{A_{sig}}^2}{A_{sig}^2}+\frac{\sigma_{A_{bkg}}^2}{A_{bkg}^2}}
\label{eq:stdSNR}
\end{equation}

\end{appendices}

\bibliographystyle{JHEP}
\bibliography{biblio}
\end{document}